\documentclass[12pt,a4paper]{hiopart}
\expandafter\let\csname equation*\endcsname\relax
\expandafter\let\csname endequation*\endcsname\relax
\usepackage{amsmath,amssymb}
\usepackage{graphicx}
\usepackage{booktabs}
\usepackage{array}
\usepackage{bm}
\usepackage{url}
\usepackage[hidelinks]{hyperref}
\AtBeginDocument{\providecommand{\texorpdfstring}[2]{#1}}
\usepackage[superscript]{cite}

\usepackage{etoolbox}
\makeatletter\patchcmd{\thebibliography}{\hfil[\arabic{enumi}]}{\hfil\arabic{enumi})}{}{\PackageWarning{main}{bibliography label patch failed}}\makeatother
\makeatletter
\def\@citess#1{\mbox{$\m@th^{\hbox{\OverciteFont{#1)}}}$}}
\renewcommand\@biblabel[1]{#1)}
\makeatother

\newcommand{\Ar}{A_\mathrm{R}}
\newcommand{\Ef}{E_\mathrm{F}}
\newcommand{\kb}{k_\mathrm{B}}
\newcommand{\Nc}{N_\mathrm{C}}
\newcommand{\epss}{\varepsilon_\mathrm{s}}
\newcommand{\phib}{\Phi_\mathrm{B}}
\newcommand{\Jte}{J_\mathrm{TE}}
\newcommand{\Jfn}{J_\mathrm{FN}}
\newcommand{\Jfe}{J_\mathrm{FE}}
\newcommand{\Jtfe}{J_\mathrm{TFE}}
\newcommand{\mdos}{m_\mathrm{dos}}
\newcommand{\mtun}{m_\mathrm{tun}}
\newcommand{\Fc}{F_\mathrm{c}}
\newcommand{\degC}{\,$^\circ$C}
\renewcommand{\figurename}{Fig.}
\makeatletter\renewcommand\thetable{\@Roman\c@table}\makeatother

\begin{document}

\title[Emission regimes and energy scales in reverse-biased 4H-SiC Schottky barriers]{Emission regimes and characteristic energy scales in reverse-biased 4H-SiC Schottky barriers: unified field-emission formulas and energy-dependent tunneling effective mass}

\author{Tetsuo Hatakeyama}
\address{Department of Electrical and Electronic Engineering, Faculty of Engineering, Toyama Prefectural University, 5180 Kurokawa, Imizu, Toyama 939-0398, Japan}
\ead{t.hatakeyama@pu-toyama.ac.jp}

\begin{abstract}
Reverse leakage in 4H-SiC Schottky barrier diodes is analyzed using a current integral with Fermi--Dirac supply and WKB transmission. The ratio of the interface field to a critical field $\Fc$ sets the depth of the thermionic-field-emission (TFE) saddle point. The Airy energy distinguishes the thermionic--tunneling crossover, the validity of the TFE formula, and the TFE/field-emission (FE) boundary, which generalizes to $c_1\kb T\approx1$ for arbitrary barrier shapes. Closed-form FE and TFE expressions are given for triangular and quadratic barriers; for the latter, the Padovani--Stratton formulas give 1.4--7.9 times (FE, $10^{19}\,\mathrm{cm^{-3}}$) and 2.5--3.6 times (TFE, low-doped layers, 300--433~K) the WKB current, whereas the present expressions agree within 15\% and 2\%. Applied to Ti/4H-SiC data (30--160\degC, 50--950~V), an energy-dependent tunneling mass from the complex band structure reduces the residual from 0.17 to 0.11--0.12~dex; it can be implemented in device simulation via the Franz two-band dispersion.
\end{abstract}


\submitto{\JJAP}
\maketitle

\section{Introduction}
\label{sec:intro}
4H-SiC has a breakdown field about ten times that of Si, which enables Schottky barrier diodes (SBDs) with higher blocking voltage and lower loss than Si SBDs\cite{Kimoto2015,Bhatnagar1992}.
To exploit this high breakdown field, a SiC SBD is designed so that the electric field at the Schottky interface under reverse bias is about ten times that of a Si SBD\cite{Hatakeyama2003}.
This is the origin of its high performance, but raising the blocking voltage and lowering the loss inevitably increases the interface field.
Unlike the breakdown field, however, the reverse characteristics of a Schottky interface are determined not by the semiconductor material but by the barrier height $\phib$ and the interface field.
The high interface field therefore increases the reverse leakage current, and the measured leakage of SiC SBDs greatly exceeds that predicted by the classical Schottky theory.
Crofton and Sriram\cite{CroftonSriram1996} calculated this leakage for 6H-SiC by numerically integrating the Wentzel--Kramers--Brillouin (WKB) transmission probability and attributed it to tunneling induced by the high interface field, and Treu et al.\cite{Treu2001} showed that, in the usual device operating temperature range, the reverse characteristics of 4H-SiC SBDs with Ti, W, and Ta electrodes are described by the Padovani--Stratton thermionic-field-emission formula with the effective mass as a fitting parameter.
Reducing the leakage by suppressing the interface field requires optimization of device structures such as the junction barrier Schottky (JBS) diode\cite{Baliga2008}, originally proposed as the pinch rectifier\cite{Baliga1984}, in which a p--n junction grid beneath the Schottky contact shields the interface under reverse bias, as well as of the barrier height and interface field\cite{Hatakeyama2003}, and such design requires a model of the Schottky-barrier leakage current.

The general treatment of current mechanisms in Schottky contacts is summarized in Refs.~\citen{SzeNg2007} and \citen{Rhoderick1988}. In reverse-biased 4H-SiC SBDs, tunneling transport contributes substantially to the reverse leakage current\cite{HatakeyamaShinohe2002,Hatakeyama2003}. Understanding the voltage and temperature dependence of the leakage requires a clear picture of how thermionic emission (TE), thermionic-field emission (TFE), and field emission (FE) are related, and of the range of validity of the approximate formulas that describe each of them.

In the theory of electron emission, following the field-emission theory of Fowler and Nordheim\cite{FowlerNordheim1928}, Murphy and Good\cite{MurphyGood1956} treated thermionic emission, field emission, and the intermediate regime from metals using the integral of the electron supply and the barrier transmission probability. For metal/semiconductor Schottky barriers, Padovani and Stratton (hereafter PS)\cite{PadovaniStratton1966}, building on Stratton's theory of field emission from semiconductors\cite{Stratton1962}, derived analytical FE and TFE formulas for forward and reverse bias together with their conditions of validity. Crowell and Rideout\cite{CrowellRideout1969} normalized the current characteristics of barriers in uniformly doped layers and examined the energy distribution of emitted electrons and the conditions for transitions between emission mechanisms. These studies connect the shift of the energy of current-carrying electrons with temperature and bias to the conditions under which each analytical formula holds.

For 4H-SiC SBDs, the present author and coworkers used triangular-barrier FE and TFE formulas to classify the dominant reverse-leakage mechanism as a function of the interface field and barrier height\cite{HatakeyamaShinohe2002}, and analyzed the reverse characteristics of Ti/4H-SiC SBDs with a TFE formula using a constant tunneling effective mass\cite{Hatakeyama2003}. More recently, Li et al.\cite{Li2022} presented an analytical model that includes the image force and the effect of the lower integration limit on the low-field side, covering the transition from TE to TFE. Hara et al.\cite{Hara2022} studied the TFE--FE transition in heavily doped SiC from the energy distribution of emitted electrons and showed that the field distribution inside the barrier must be taken into account to describe the transition.

These studies indicate that the analysis of reverse transport requires relating three quantities: the characteristic energy describing the barrier curvature, the electric field in the tunneling region, and the energy of the current-carrying electrons. In low-doped layers under high reverse bias, the tunneling path of the electrons that dominate the current is confined to the vicinity of the interface, and if the field changes little over that path, the barrier can be approximated as triangular. Making explicit how this approximation relates to the conventional analysis of the quadratic barrier, and which scales organize the TE, TFE, and FE regimes, provides a clear basis for choosing analytical formulas and stating their validity. Murakami et al.\cite{Murakami2023} analyzed the reverse tunneling current on the basis of the complex band structure of 4H-SiC. When the tunneling energy lies deep in the gap, the validity of approximating the in-gap dispersion by a constant mass must also be examined.

In this paper, we examine reverse transport in 4H-SiC Schottky barriers theoretically, using a current integral based on Fermi--Dirac supply and WKB transmission. We first use the critical field $\Fc$ and the Airy energy and length to relate the representative tunneling energy to the emission mechanism; after showing from exact solutions of the Airy equation that these scales govern the decay of the electron wave function in the barrier, we distinguish the conditions for crossover between current components from the conditions of validity of the approximate formulas. We also extend the TFE/FE boundary to general barrier shapes and relate it to the transition condition of Hara et al.\cite{Hara2022}. We then give closed-form expressions for the reverse FE and TFE currents of triangular and quadratic barriers and re-examine the PS reverse FE and TFE formulas. Next, we introduce the image force and an energy-dependent tunneling effective mass and examine their effect on the field and temperature dependence of the current. As an application, we reanalyze measured data for the Ti/4H-SiC SBD of Ref.~\citen{Hatakeyama2003}, including temperatures not previously reported, and compare the reproducibility of each model and its sensitivity to the sample parameters. Finally, we show how the energy-dependent mass can be implemented in a device simulator. The main derivations and numerical methods are given in the appendices, and detailed derivations of each current formula are given in the supplementary data.

\section{Theoretical framework}
\label{sec:model}

\subsection{Energy reference and band diagrams}
\begin{figure}[htbp]
 \centering
 \includegraphics[width=0.8\linewidth]{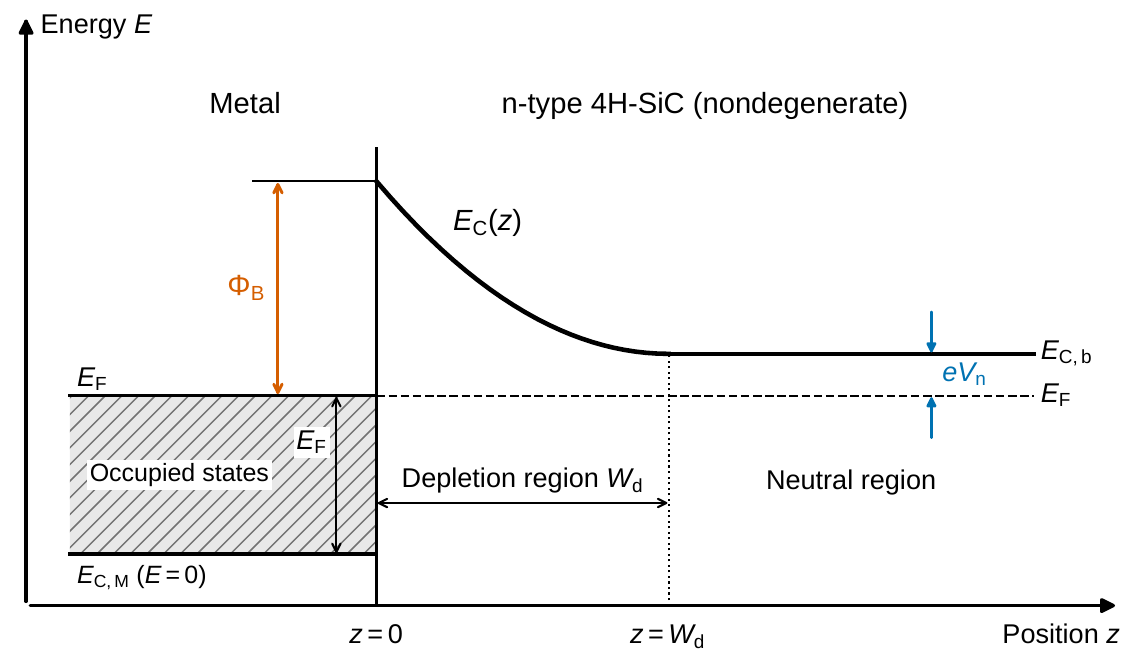}
 \caption{Band diagram and energy reference of a metal/n-type 4H-SiC Schottky junction in thermal equilibrium.
 The bottom of the metal supply band, $E_{\mathrm{C,M}}$, which lies below the Fermi level $\Ef$, is the energy origin.
 The barrier top is $E_0=\Ef+\phib$, and the conduction-band edge in the depletion layer varies quadratically. The hatched region shows the occupied states of the metal.}
 \label{fig:band}
\end{figure}

\begin{figure}[htbp]
 \centering
 \includegraphics[width=0.85\linewidth]{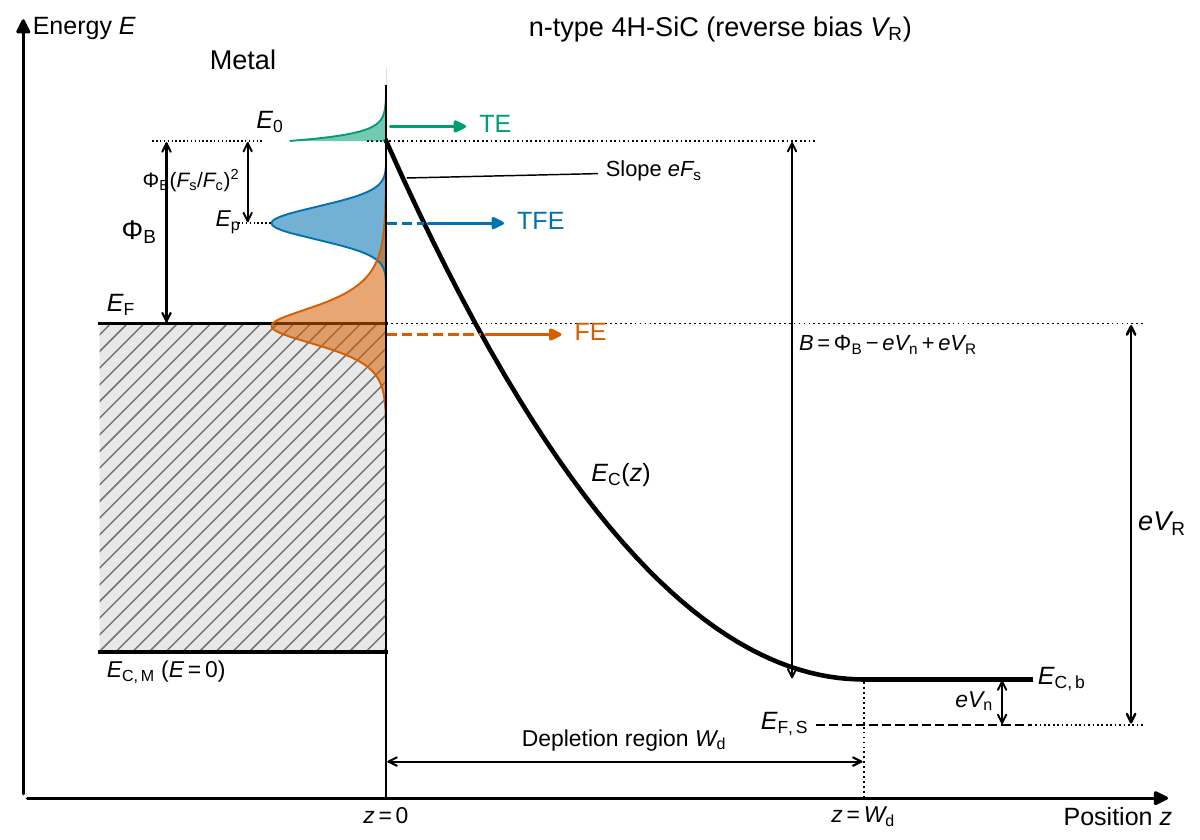}
 \caption{Schematic band diagram under reverse bias $V_\mathrm{R}$. The metal Fermi level $\Ef$ and the bulk Fermi level of the semiconductor $E_{\mathrm{F,S}}$ differ by $eV_\mathrm{R}$,
 and the band bending is $B=\phib-eV_\mathrm{n}+eV_\mathrm{R}$. For clarity, $eV_\mathrm{R}$ is drawn as a few eV; in the experiment $V_\mathrm{R}$ reaches about 1000~V and $B\gg\phib$.
 The filled areas schematically show the energy distribution of electrons injected from the metal, $\mathcal{T}(E)\ln[1+e^{-(E-\Ef)/\kb T}]$, for the three paths:
 TE (green, thermal tail above the barrier top), TFE (blue, nearly Gaussian distribution centered at the saddle point $E_\mathrm{p}$ inside the barrier), and
 FE (orange, distribution concentrated near the Fermi level).
 The three distributions correspond to different field ranges and are drawn together in one schematic diagram; they do not occur at the same field.
 Above the Fermi level, the FE distribution decays with the width $\kb T/(1-p)$ ($p=\Fc/F_\mathrm{s}$); below it, the supply approaches $(\Ef-E)/\kb T$, and $\kb T/p$ is the decay energy of the exponential factor (supplementary data, Sect.~S3). The FE distribution is drawn for $F_\mathrm{s}/\Fc=1.5$.
 The TFE saddle point, which exists for $F_\mathrm{s}<\Fc$, lies $\phib(F_\mathrm{s}/\Fc)^2$ below the barrier top (Sect.~\ref{sec:index}). Dashed lines indicate the classically forbidden region traversed by tunneling.}
 \label{fig:band_rev}
\end{figure}

Figures~\ref{fig:band} and \ref{fig:band_rev} show the band diagrams of a metal/n-type 4H-SiC Schottky junction in thermal equilibrium and under reverse bias, respectively.
The electron supply of the metal is described by a parabolic band, and its band bottom $E_{\mathrm{C,M}}$ is used as the common energy origin.
With the metal Fermi level $\Ef$ and the Schottky barrier height $\phib$ (an energy), the barrier top is $E_0=\Ef+\phib$.
The difference between the conduction-band edge and the Fermi level in the semiconductor bulk is $eV_\mathrm{n}=\kb T\ln(\Nc/N_\mathrm{D})$\cite{SzeNg2007}.
Under reverse bias $V_\mathrm{R}\geq0$, the Fermi level of the semiconductor drops to $E_{\mathrm{F,S}}=\Ef-eV_\mathrm{R}$, and the band bending increases to
\begin{equation}
 B=\phib-eV_\mathrm{n}+eV_\mathrm{R}.
 \label{eq:B}
\end{equation}
Because the barrier height $\phib$ does not depend on bias (apart from the image force), the interface field $F_\mathrm{s}$ increases and the barrier becomes thinner.
Electrons injected from the metal into the semiconductor either pass over the barrier top (TE), tunnel after thermal excitation to an intermediate energy (TFE), or tunnel from near the Fermi level (FE) (Fig.~\ref{fig:band_rev}).
The reverse leakage current is this flow of electrons from the metal into the semiconductor. Below, $z$ denotes the direction normal to the interface, and the metal occupies $z<0$.

\subsection{Electron supply and reverse current integral}
The one-way current density from the metal into the semiconductor, after integration over the momentum parallel to the interface, is
\begin{equation}
 J=\frac{\Ar T}{\kb}\int_0^{\infty}\mathcal{T}(E)\,
 \ln\!\left[1+\exp\!\left(-\frac{E-\Ef}{\kb T}\right)\right]dE,
 \qquad
 \Ar=\frac{4\pi e g_\mathrm{v}\mdos\kb^2}{h^3}
 \label{eq:J}
\end{equation}
(supplementary data, Sects.~S1 and S2)\cite{TsuEsaki1973}, where $E$ is the kinetic energy normal to the interface and $\mathcal{T}(E)$ is the transmission probability.
For 4H-SiC, the valley degeneracy $g_\mathrm{v}=3$ and the density-of-states mass parallel to the interface $\mdos=\sqrt{0.57\times0.28}\,m_0$\cite{Wellenhofer1997} give $\Ar=144\,\mathrm{A\,cm^{-2}K^{-2}}$.
Above the barrier top we set $\mathcal{T}=1$; this part gives the TE current $\Jte=\Ar T^2\exp(-\phib/\kb T)$.
The lower limit of the integral is 0 on the common energy scale, or $-\Ef$ when the energy is measured from the Fermi level.

We note explicitly that $\Ar$, determined by the semiconductor mass, is used also for the electron flow from the metal into the semiconductor.
In thermal equilibrium, the electron flows from the metal into the semiconductor and from the semiconductor into the metal are equal by detailed balance.
The latter is the flux of conduction-band electrons in the semiconductor and is given by $\Ar T^2\exp(-\phib/\kb T)$, determined by the density-of-states mass of the semiconductor.
The former must therefore be expressed with the same $\Ar$.
Here $\mdos=\sqrt{m_xm_y}$ is the in-plane density-of-states mass of one valley, with $m_x$ and $m_y$ the effective masses parallel to the interface.
Crowell\cite{Crowell1965,Crowell1969} showed that, when the wave vector parallel to the interface is conserved, only electrons incident from the metal within an ``acceptance cone'' determined by the energy--wave-vector relation of the semiconductor can enter the semiconductor;
hence the Richardson constant is determined by the semiconductor's in-plane density-of-states mass $\mdos$, not by the metal, and is the same for forward and reverse bias.
He further showed that the mass in the prefactor of the TFE current is this same mass as in TE and differs from the tunneling effective mass along the current direction\cite{Crowell1969}.
This is why we distinguish the supply mass $\mdos$ from the tunneling mass $\mtun$.
In the tunneling integral we likewise assume that the parallel wave vector is conserved and that the transmission depends only on the energy $E$ of the motion normal to the interface, so that every transverse mode sees the same $\mathcal{T}(E)$ and the same $\Ar$ applies.
Corrections to $\Ar$ due to quantum-mechanical reflection and phonon scattering\cite{CrowellSze1966} are not considered here.
In contrast, Nicholls et al.\cite{Nicholls2019} presented an analysis of 4H-SiC SBDs in which the metal mass determines the current, so that views differ on this point.
For $eV_\mathrm{R}\gg\kb T$ under reverse bias, the current from the semiconductor into the metal is negligible.

\subsection{WKB transmission probability and reference barrier models}
The WKB transmission probability is $\mathcal{T}(E)=\exp[-G(E)]$ with
\begin{equation}
 G(E)=\frac{2}{\hbar}\int_{z_1}^{z_2}\sqrt{2\mtun\,[U(z)-E]}\,dz,
 \label{eq:G}
\end{equation}
where $z_1$ and $z_2$ are the ends of the classically forbidden region (the interface and the turning point on the semiconductor side without the image force, and the two turning points with it). We consider the following barriers.

\paragraph{Triangular barrier.}
For $U(z)=E_0-eF_\mathrm{s}z$ with a constant interface field $F_\mathrm{s}$,
\begin{equation}
 G_\mathrm{tri}(E)=\frac{4\sqrt{2\mtun}}{3e\hbar F_\mathrm{s}}(E_0-E)^{3/2}.
 \label{eq:Gtri}
\end{equation}

\paragraph{Quadratic barrier.}
For $U(z)=E_{\mathrm{C,b}}+B(1-z/W_\mathrm{d})^2$ in the depletion approximation, with $u=(E-E_{\mathrm{C,b}})/B$ ($0<u<1$, i.e., $E_{\mathrm{C,b}}<E<E_0$),
\begin{equation}
 G_\mathrm{quad}(E)=\frac{B}{E_{00}}\left[\sqrt{1-u}-u\,\mathrm{arcosh}\frac{1}{\sqrt u}\right],
 \qquad
 E_{00}=\frac{e\hbar}{2}\sqrt{\frac{N_\mathrm{D}}{\mtun\epss}}
 \label{eq:Gquad}
\end{equation}
(supplementary data, Sect.~S4). $E_{00}$ is the only energy, $\frac\hbar2\sqrt{k/\mtun}$, that can be formed from the barrier curvature $k=d^2U/dz^2=e^2N_\mathrm{D}/\epss$, the mass, and $\hbar$,
and it is the unit that makes the Schr\"odinger equation for the quadratic potential dimensionless (Sect.~\ref{sec:scales}).
Below the bulk conduction-band edge ($E<E_{\mathrm{C,b}}$) there are no states in the semiconductor, and we set $\mathcal{T}=0$ in the numerical integration for the quadratic barrier.
The triangular barrier is the tangent of the quadratic barrier at the interface.

\subsection{Interface field and punch-through}
In the depletion approximation, the interface field is
\begin{equation}
 F_\mathrm{s}=\begin{cases}
 \sqrt{2eN_\mathrm{D}(V_\mathrm{R}+V_\mathrm{bi})/\epss}, & W_\mathrm{d}\leq d,\\[2pt]
 (V_\mathrm{R}+V_\mathrm{bi})/d+eN_\mathrm{D}d/(2\epss), & W_\mathrm{d}>d,
 \end{cases}
 \qquad eV_\mathrm{bi}=\phib-eV_\mathrm{n},
 \label{eq:Fs}
\end{equation}
where $d$ is the drift-layer thickness and $\epss=10.03\varepsilon_0$ is the permittivity along the c-axis\cite{PatrickChoyke1970}. Here $W_\mathrm{d}=\sqrt{2\epss(V_\mathrm{R}+V_\mathrm{bi})/(eN_\mathrm{D})}$ is the depletion width for a semi-infinite uniformly doped layer, and $W_\mathrm{d}>d$ is used as the criterion for punch-through. The second line corresponds to a fully depleted drift layer (Appendix~C).

\section{Emission regimes and characteristic energy scales}
\label{sec:index}
This section expresses the boundaries between emission regimes as comparisons of three energies: a quantum energy set by the slope of the barrier wall, the thermal energy $\kb T$, and the barrier height $\phib$.
Section~\ref{sec:emission} shows the distribution of emitted electrons and the critical field $\Fc$, and Sect.~\ref{sec:scales} introduces the characteristic energy scales $E_\mathrm{A}$ and $E_{00}$.
Section~\ref{sec:airy} expresses the boundaries with these scales, and Sect.~\ref{sec:closed} gives closed-form expressions for the reverse FE and TFE currents. Section~\ref{sec:approx} examines the analytical formulas and the validity of the approximations, and Sect.~\ref{sec:prev} re-examines previously reported closed-form expressions and transition conditions.

\subsection{Distribution of emitted electrons and the critical field}
\label{sec:emission}
\begin{figure}[htbp]
 \centering
 \includegraphics[width=\linewidth]{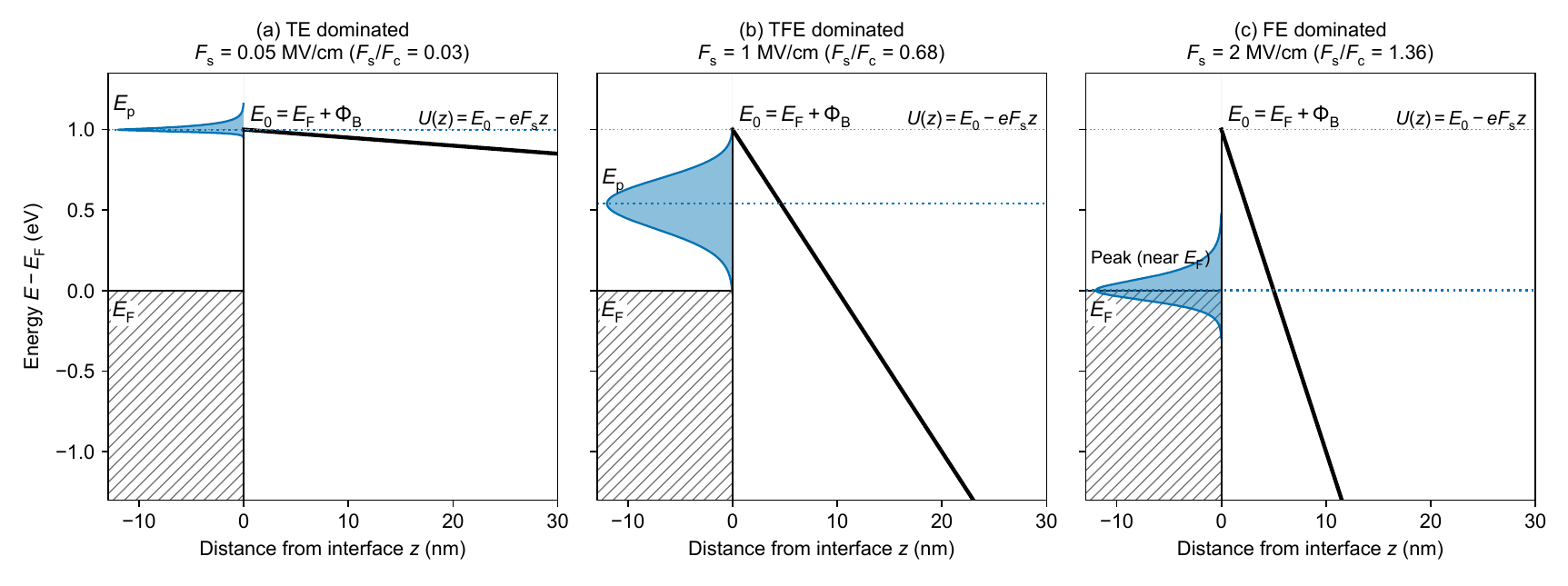}
 \caption{Energy distribution of electrons injected from the metal through a triangular barrier,
 $\mathcal{T}(E)\ln[1+e^{-(E-\Ef)/\kb T}]$ (blue fill, normalized to unity at its maximum in each panel; the blue dotted line marks the peak).
 (a) $F_\mathrm{s}=0.05\,\mathrm{MV/cm}$ (TE dominated),
 (b) $1.0\,\mathrm{MV/cm}$ (TFE dominated; the peak is at $E_\mathrm{p}$ of Eq.~(\ref{eq:Ep})),
 (c) $2.0\,\mathrm{MV/cm}$ (FE dominated).
 $T=300\,\mathrm K$, $\phib=1.00\,\mathrm{eV}$, $\mtun=0.31m_0$, $\Fc=1.475\,\mathrm{MV/cm}$.}
 \label{fig:emission}
\end{figure}

The integrand of Eq.~(\ref{eq:J}), $\mathcal{T}(E)\ln[1+e^{-(E-\Ef)/\kb T}]$, is the energy distribution of injected electrons.
In the Boltzmann approximation, the energy that minimizes the sum of the triangular-barrier exponent~(\ref{eq:Gtri}) and $(E-\Ef)/\kb T$ (the saddle point) is
\begin{equation}
 E_\mathrm{p}-\Ef=\phib\left[1-\left(\frac{F_\mathrm{s}}{\Fc}\right)^2\right],\qquad
 \Fc=\frac{2\kb T\sqrt{2\mtun\phib}}{e\hbar}
 \label{eq:Ep}
\end{equation}
(Appendix~A). The saddle point thus lies at the depth $\Delta\equiv E_0-E_\mathrm{p}=\phib(F_\mathrm{s}/\Fc)^2$ below the barrier top $E_0=\Ef+\phib$; we call $\Delta$ the saddle depth. For 300~K, $\phib=1.00\,\mathrm{eV}$, and $\mtun=0.31m_0$, $\Fc=1.475\,\mathrm{MV/cm}$.
Figure~\ref{fig:emission} shows the energy distribution of emitted electrons for the triangular barrier.
For $F_\mathrm{s}\ll\Fc$, the distribution is concentrated at the barrier top (TE);
for $F_\mathrm{s}<\Fc$, electrons tunnel at $E_\mathrm{p}$ inside the barrier (TFE);
and for $F_\mathrm{s}>\Fc$, the saddle point moves below the Fermi level (FE).
$\Fc$ is also the formal boundary between the saddle-point approximation for TFE and the linearized approximation for FE (Appendix~A).

\subsection{Characteristic energy scales \texorpdfstring{$E_\mathrm{A}$}{EA} and \texorpdfstring{$E_{00}$}{E00}}
\label{sec:scales}
The only quantity that characterizes the triangular barrier $U=E_0-eF_\mathrm{s}z$ is the wall slope $eF_\mathrm{s}$; the only energy and length that can be formed from it, the mass, and $\hbar$ are
\begin{equation}
 E_\mathrm{A}=\left(\frac{(e\hbar F_\mathrm{s})^2}{2\mtun}\right)^{1/3},\qquad l_\mathrm{A}=\frac{E_\mathrm{A}}{eF_\mathrm{s}}=\left(\frac{\hbar^2}{2\mtun eF_\mathrm{s}}\right)^{1/3}.
 \label{eq:EA}
\end{equation}
Measuring length in units of $l_\mathrm{A}$ and energy in units of $E_\mathrm{A}$, the Schr\"odinger equation becomes the parameter-free Airy equation $d^2\psi/d\xi^2=\xi\psi$ in $\xi=(E_0-E-eF_\mathrm{s}z)/E_\mathrm{A}$\cite{DLMF}.
We call $E_\mathrm{A}$ the Airy energy and $l_\mathrm{A}$ the Airy length.
For an electron at a depth $E_0-E$ below the barrier top, the exponent $G_\mathrm{tri}=\frac43[(E_0-E)/E_\mathrm{A}]^{3/2}$, the tunneling distance $l_\mathrm{A}(E_0-E)/E_\mathrm{A}$, and the decay length at the interface $l_\mathrm{A}\sqrt{E_\mathrm{A}/(E_0-E)}$ are all determined by the ratio $(E_0-E)/E_\mathrm{A}$ alone.
$E_\mathrm{A}$ is an energy scale; it is not the energy of the electron.
Electrons in the triangular well of a MOS inversion layer\cite{Stern1972,Ando1982} obey the same Airy equation; only the boundary conditions differ (supplementary data, Sect.~S6.3).

Similarly, the only energy and length that can be formed from the curvature $k$ of the quadratic barrier $U=E_{\mathrm{C,b}}+\frac12k(W_\mathrm{d}-z)^2$ ($k=e^2N_\mathrm{D}/\epss$), the mass, and $\hbar$ are
\begin{equation}
 E_{00}=\frac\hbar2\sqrt{\frac{k}{\mtun}}=\frac{e\hbar}{2}\sqrt{\frac{N_\mathrm{D}}{\mtun\epss}},\qquad l_{00}=\left(\frac{\hbar^2}{\mtun k}\right)^{1/4},
 \label{eq:E00l00}
\end{equation}
where $E_{00}$ is the same as in Eq.~(\ref{eq:Gquad}). With $\eta=(W_\mathrm{d}-z)/l_{00}$ and $\varepsilon=E-E_{\mathrm{C,b}}$, the Schr\"odinger equation becomes
\begin{equation}
 \frac{d^2\psi}{d\eta^2}=\left(\eta^2-\frac{\varepsilon}{E_{00}}\right)\psi
 \label{eq:weber}
\end{equation}
(Weber's equation\cite{DLMF12}). This is the equation of a harmonic oscillator; with boundary conditions requiring decay on both sides, $E_{00}$ corresponds to the zero-point energy.
In a Schottky barrier, the potential is truncated at the interface and the electrons are not bound, so $E_{00}$ and $l_{00}$ appear simply as the units of this equation (supplementary data, Sect.~S6.1).

The slope of a quadratic wall varies with height; the local Airy energy at a height $\varepsilon$ above the bottom of the well is $E_\mathrm{A}(\varepsilon)=(4\varepsilon E_{00}^2)^{1/3}$.
Its value at the barrier top ($\varepsilon=B$) is the interface Airy energy,
\begin{equation}
 E_\mathrm{A}=\left(4BE_{00}^2\right)^{1/3}
 \label{eq:EA_E00}
\end{equation}
(supplementary data, Sect.~S6.2). From Eq.~(\ref{eq:EA}) and the definition of $\Fc$ in Eq.~(\ref{eq:Ep}), $F_\mathrm{s}/\Fc=E_\mathrm{A}^{3/2}/(2\kb T\sqrt{\phib})$; substituting Eq.~(\ref{eq:EA_E00}) gives
\begin{equation}
 \boxed{\frac{F_\mathrm{s}}{\Fc}=\frac{E_{00}}{\kb T}\sqrt{\frac{B}{\phib}}}.
 \label{eq:index}
\end{equation}

Equation~(\ref{eq:index}) leads to the following four points.
\begin{enumerate}
 \item[(i)] \textbf{At zero bias it reduces to the PS classification.}
 For $eV_\mathrm{n}\ll\phib$, $B\simeq\phib$ and $F_\mathrm{s}/\Fc\simeq E_{00}/\kb T$, so the TFE/FE boundary $E_{00}\approx\kb T$ of PS\cite{PadovaniStratton1966} corresponds to $F_\mathrm{s}=\Fc$ at zero bias.
 \item[(ii)] \textbf{Under reverse bias the change is carried by $\sqrt{B/\phib}$.}
 $E_{00}/\kb T$ is determined by the curvature alone and does not depend on bias. The tunneling electrons, however, are near the interface ($\varepsilon\approx B$), where the wall slope $eF_\mathrm{s}=\sqrt{2kB}$ increases as the well deepens under reverse bias.
 This effect is the factor $\sqrt{B/\phib}$, and Eq.~(\ref{eq:index}) extends the PS index to high bias.
 \item[(iii)] \textbf{Even in low-doped layers, tunneling becomes important at high reverse bias.}
 For $N_\mathrm{D}=5\times10^{15}\,\mathrm{cm^{-3}}$, $E_{00}=0.74\,\mathrm{meV}$ (about $\kb T/35$), which, in a classification by $E_{00}/\kb T$ alone, falls in the TE regime regardless of bias.
 At $B=300\,\mathrm{eV}$ (a reverse voltage of about 300~V, $F_\mathrm{s}\approx0.74\,\mathrm{MV/cm}$), however, $E_\mathrm{A}\approx0.09\,\mathrm{eV}$ (about $3\kb T$) and $l_\mathrm{A}\approx1\,\mathrm{nm}$, which is the TFE regime.
 \item[(iv)] \textbf{In low-doped layers, the saddle point of the triangular barrier coincides with that of the quadratic barrier.}
 The TFE saddle point of the quadratic barrier lies $B\tanh^2(E_{00}/\kb T)$ below the barrier top (Appendix~B). For $E_{00}\ll\kb T$ this becomes $B(E_{00}/\kb T)^2=\phib(F_\mathrm{s}/\Fc)^2$, the saddle depth of the triangular barrier in Eq.~(\ref{eq:Ep}).
\end{enumerate}
Equation~(\ref{eq:index}) assumes the depletion approximation $F_\mathrm{s}^2=2N_\mathrm{D}B/\epss$ ($B$ in J) and the constant-mass $\Fc$. After punch-through, $F_\mathrm{s}$ is obtained directly from Eq.~(\ref{eq:Fs}), and $F_\mathrm{s}/\Fc$ and $E_\mathrm{A}$ are calculated from it.

\subsection{Boundaries between emission regimes}
\label{sec:airy}
For the triangular barrier in the Boltzmann approximation, the saddle point is where the slope of the exponent with respect to depth, $2\Delta^{1/2}/E_\mathrm{A}^{3/2}$, equals the slope of the supply, $1/\kb T$. In terms of $E_\mathrm{A}$, the saddle depth $\Delta$ (Sect.~\ref{sec:emission}) is
\begin{equation}
 \Delta=\phib\left(\frac{F_\mathrm{s}}{\Fc}\right)^2=\frac{E_\mathrm{A}^3}{4(\kb T)^2}.
 \label{eq:Delta_EA}
\end{equation}
The depth of the current-carrying electrons is determined only by $E_\mathrm{A}$ and $\kb T$, and $F_\mathrm{s}=\Fc$ is the point at which this depth reaches $\phib$.

\paragraph{Competition between TE and tunneling.}
For a triangular barrier with constant mass and without image force, and as long as the Boltzmann supply applies, the competition between the TE and tunneling components is determined by $x\equiv E_\mathrm{A}/\kb T$ alone.
Hatakeyama et al.\cite{HatakeyamaShinohe2002} obtained the TFE current of the triangular barrier by the saddle-point method (the Hatakeyama formula; derived in Appendix~A). In terms of $x$ it can be written as
\begin{equation}
 \Jtfe=\Jte\cdot\sqrt\pi\,x^{3/2}\exp\left(\frac{x^3}{12}\right),
 \label{eq:tfe_x}
\end{equation}
so that the ratio of the TFE and TE currents contains neither the barrier height nor the donor concentration. The exponent $x^3/12$ equals $\Delta/(3\kb T)$.
Even for the WKB integral without the saddle-point approximation, for $\phib\gg E_\mathrm{A}$,
\begin{equation}
 \frac{J_\mathrm{tun}}{\Jte}=f(x)=\int_0^\infty\exp\left[s-\frac43\left(\frac{s}{x}\right)^{3/2}\right]ds
 \label{eq:fx}
\end{equation}
($s=(E_0-E)/\kb T$) is again a universal function of $x$ alone.
$f(x)$ increases monotonically with $x$ and equals unity at $x\approx0.80$. The TE/TFE boundary, where the TE and tunneling currents are equal, is therefore $E_\mathrm{A}\approx0.80\,\kb T$, which corresponds to $F_\mathrm{s}\approx0.085\,\mathrm{MV/cm}$ at 300~K for $\mtun=0.31m_0$. At higher fields the tunneling current exceeds the TE current.
Equation~(\ref{eq:tfe_x}) is the saddle-point approximation of $f(x)$ and becomes more accurate as the saddle point lies deeper below the barrier top. We take $\Delta=3\kb T$ ($x=12^{1/3}\approx2.3$) as the guide for the TFE saddle-point approximation; for $x\ge2$ it is within 13\% of $f(x)$ (supplementary data, Sect.~S6.5).

\paragraph{Boundary between TFE and FE.}
For a general barrier shape, the TFE/FE boundary is expressed by the slope of the exponent at the Fermi level,
\begin{equation}
 c_1\equiv-\left.\frac{\partial G}{\partial E}\right|_{E=\Ef}.
 \label{eq:c1}
\end{equation}
$c_1$ is the coefficient of the PS FE formula\cite{PadovaniStratton1966}.
Expanding the exponent to first order near the Fermi level, $G(E)\simeq b_1-c_1(E-\Ef)$, the transmission probability increases as $e^{-b_1}e^{c_1(E-\Ef)}$, whereas the supply decreases as $e^{-(E-\Ef)/\kb T}$ for $E-\Ef\gg\kb T$ (Boltzmann approximation). In this approximation, their product, the integrand of Eq.~(\ref{eq:J}), is
\begin{equation}
 \mathcal{T}(E)\ln\left[1+e^{-(E-\Ef)/\kb T}\right]\simeq e^{-b_1}\exp\left[(c_1\kb T-1)\,\frac{E-\Ef}{\kb T}\right],
 \label{eq:c1_balance}
\end{equation}
and the sign of its slope is set only by whether $c_1\kb T$ is larger or smaller than unity.
For $c_1\kb T>1$ the integrand keeps increasing above the Fermi level, and the main contribution moves up to the saddle point, where the curvature of the exponent takes over (TFE).
For $c_1\kb T<1$ the Boltzmann tail decreases with energy, and the current comes from near the Fermi level (FE).
Thus $c_1\kb T=1$ is the point at which the Boltzmann saddle reaches the metal Fermi level; it is also the convergence limit of the FE integral with the linearized exponent (Sect.~\ref{sec:closed}), and we refer to it as the formal TFE/FE boundary.
With Fermi--Dirac supply, the crossover is gradual and its location depends on the criterion used to define it: within the linear expansion, the peak of the integrand lies at $\Ef$ for $c_1\kb T=1/(2\ln2)=0.72$ and at $\Ef+\kb T$ for $c_1\kb T=0.86$ (supplementary data, Sect.~S6.6).
Crofton and Sriram\cite{CroftonSriram1996} also stated the FE/TFE boundary as a comparison of $1/c_1$ with $\kb T$.

For the triangular barrier,
\begin{equation}
 c_1=\frac{2\sqrt{\phib}}{E_\mathrm{A}^{3/2}},\qquad
 c_1\kb T=\frac{2\kb T\sqrt{\phib}}{E_\mathrm{A}^{3/2}}=\frac{\Fc}{F_\mathrm{s}},
 \label{eq:c1kT_tri}
\end{equation}
so that $F_\mathrm{s}=\Fc$ is the same condition as $c_1\kb T=1$.
The energy $1/c_1=(E_\mathrm{A}/2)\sqrt{E_\mathrm{A}/\phib}$ is the interval over which the transmission probability at the Fermi level changes by a factor of $e$;
the TFE/FE boundary compares this quantum energy, set by $E_\mathrm{A}$ and $\phib$, with $\kb T$.

For the quadratic barrier (supplementary data, Sect.~S4.4),
\begin{equation}
 c_1=\frac{1}{E_{00}}\operatorname{artanh}\rho=\frac{2\sqrt{\phib}}{E_\mathrm{A}^{3/2}}\,\frac{\operatorname{artanh}\rho}{\rho},\qquad
 c_1\kb T=\frac{\Fc}{F_\mathrm{s}}\,\frac{\operatorname{artanh}\rho}{\rho},\qquad \rho=\sqrt{\frac{\phib}{B}},
 \label{eq:c1_quad}
\end{equation}
where $E_\mathrm{A}$ and $F_\mathrm{s}$ are the interface values. This expression holds for $X=eV_\mathrm{R}-eV_\mathrm{n}>0$, i.e., $\rho<1$, when the metal Fermi level lies above the bulk conduction-band edge.
The factor $\operatorname{artanh}\rho/\rho=1+\rho^2/3+\cdots\ge1$ is the correction for the curvature of the barrier along the tunneling path and approaches unity as the band bending $B$ increases.
For a high-voltage SBD with a low-doped drift layer under reverse bias\cite{Hatakeyama2003} (the sample of Sect.~\ref{sec:exp}, $B\gtrsim50\,\mathrm{eV}$), the factor is below 1.007, i.e., practically unity.
For Schottky barriers formed on heavily doped epitaxial layers\cite{Hara2022}, by contrast, $B$ is small, and the factor is about 1.5 at the transition for $N_\mathrm{D}=1.2\times10^{19}\,\mathrm{cm^{-3}}$ ($B\approx2\,\mathrm{eV}$).
In heavily doped layers, therefore, the TFE/FE boundary cannot be expressed by the single field $F_\mathrm{s}=\Fc$.
The condition $c_1\kb T=1$, in contrast, gives the TFE/FE boundary independently of the barrier shape and reduces to $F_\mathrm{s}=\Fc$ in the triangular limit.

These boundaries are summarized in Table~\ref{tab:regime}.
The condition $J_\mathrm{tun}=\Jte$ and the condition at which the TFE saddle-point approximation becomes accurate ($\Delta=3\kb T$) depend only on $x$; in terms of the field, $F_\mathrm{s}\propto\sqrt{\mtun}(\kb T)^{3/2}/(e\hbar)$, independent of the barrier height and the donor concentration.
The TFE/FE boundary, by contrast, is set by whether the saddle point reaches the Fermi level, giving $\Fc\propto T\sqrt{\phib}$.

\begin{table}[htbp]
 \caption{\label{tab:regime}Boundaries between emission regimes expressed by the Airy energy $E_\mathrm{A}$. Values are for 300~K, $\phib=1.00\,\mathrm{eV}$, and $\mtun=0.31m_0$ (no image force).}
 \footnotesize
 \begin{tabular*}{\textwidth}{@{}l@{\extracolsep{\fill}}lrr}
 \br
 Boundary & Condition & $E_\mathrm{A}$ (eV) & $F_\mathrm{s}$ (MV/cm)\\
 \mr
 $J_\mathrm{tun}=\Jte$ & $f(x)=1\ \Rightarrow\ E_\mathrm{A}\approx0.80\,\kb T$ & 0.021 & 0.085\\
 TFE saddle point & $\Delta=3\kb T\ \Rightarrow\ E_\mathrm{A}=12^{1/3}\kb T\approx2.3\,\kb T$ & 0.059 & 0.41\\
 TFE/FE & $\Delta=\phib\Leftrightarrow c_1\kb T=1\Rightarrow E_\mathrm{A}=[4\phib(\kb T)^2]^{1/3}$ & 0.139 & 1.475 ($=\Fc$)\\
 \br
 \end{tabular*}
\end{table}

\begin{figure}[htbp]
 \centering
 \includegraphics[width=0.6\linewidth]{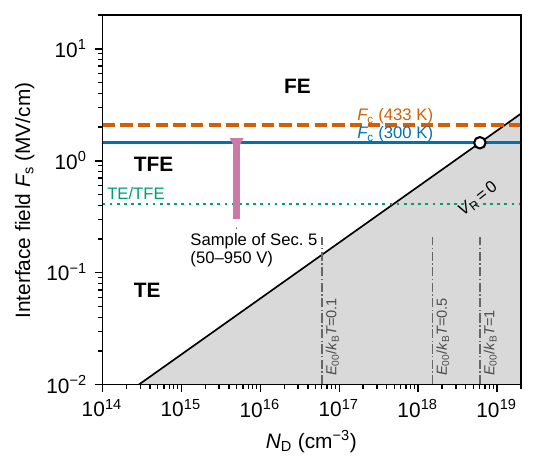}
 \caption{Emission regimes in the plane of donor concentration $N_\mathrm{D}$ and interface field $F_\mathrm{s}$ ($\phib=0.96\,\mathrm{eV}$, $\mtun=0.31m_0$).
 Solid (blue) and dashed (orange) lines: $\Fc$ at 300 and 433~K; dotted (green) line: guide for the TFE saddle-point approximation, $\Delta=3\kb T$, i.e., $F_\mathrm{s}/\Fc=\sqrt{3\kb T/\phib}$ (300~K);
 dash-dotted lines: contours of $E_{00}/\kb T$ at 300~K; gray: fields below the zero-bias value, which cannot be reached. The open circle marks the intersection of the $V_\mathrm{R}=0$ line, $\Fc$ at 300~K, and $E_{00}/\kb T=1$.
 Thick (purple) bar: measurement range of the sample whose measured characteristics are analyzed in Sect.~\ref{sec:exp}.}
 \label{fig:regime}
\end{figure}

\paragraph{Regimes in the plane of donor concentration and interface field.}
Figure~\ref{fig:regime} shows the emission regimes in the $N_\mathrm{D}$--$F_\mathrm{s}$ plane.
The $V_\mathrm{R}=0$ line is the zero-bias interface field $F_\mathrm{s}(0)=\sqrt{2eN_\mathrm{D}B/\epss}$ drawn with $B\simeq\phib$, i.e., the lowest field attainable at each $N_\mathrm{D}$.
In this approximation Eq.~(\ref{eq:index}) reduces to $F_\mathrm{s}/\Fc=E_{00}/\kb T$, so the $V_\mathrm{R}=0$ line and the $\Fc$ line intersect on the $E_{00}/\kb T=1$ contour ($N_\mathrm{D}\approx6\times10^{18}\,\mathrm{cm^{-3}}$, open circle).
In heavily doped regions such as ohmic contacts, $F_\mathrm{s}>\Fc$ already at zero bias,
whereas SiC drift layers, although $E_{00}/\kb T<0.1$, enter the TFE regime at high reverse voltage and approach $\Fc$.
The Ti/4H-SiC SBD of Ref.~\citen{Hatakeyama2003} ($N_\mathrm{D}=5\times10^{15}\,\mathrm{cm^{-3}}$, 50--950~V; hereafter ``the sample''), whose measured characteristics are analyzed in Sect.~\ref{sec:exp}, covers $F_\mathrm{s}=0.30$--$1.40\,\mathrm{MV/cm}$,
and $F_\mathrm{s}/\Fc$ reaches 0.96 at 950~V and 30\degC.

\subsection{Closed-form expressions for the reverse FE and TFE currents}
\label{sec:closed}
The WKB integral~(\ref{eq:J}) can be evaluated in closed form, i.e., without an integral, by expanding the integrand around the energy that carries most of the current.
For the FN and FE formulas, the exponent $G(E)$ is expanded to first order at the Fermi level ($b_1$ and $c_1$ of Eq.~(\ref{eq:c1_balance})) and the remaining integral is performed analytically.
For the TFE formula, the supply is approximated by the Boltzmann form, the exponent of the integrand is expanded to second order about the saddle point, and the Gaussian integral is performed.
This subsection gives the resulting expressions and an outline of their derivation for the triangular and quadratic barriers.
All refer to the one-way current from the metal into the semiconductor, with a constant tunneling mass and without the image force.
The detailed derivations are given in Sects.~S3 (triangular barrier) and S4 (quadratic barrier) of the supplementary data.

\paragraph{Triangular barrier.}
The value and slope of the exponent~(\ref{eq:Gtri}) at the Fermi level are
\begin{equation}
 b_1=\frac{4\sqrt{2\mtun}\,\phib^{3/2}}{3e\hbar F_\mathrm{s}},\qquad
 c_1=\frac{2\sqrt{2\mtun\phib}}{e\hbar F_\mathrm{s}},
 \label{eq:b1c1_tri}
\end{equation}
where $c_1$ is the same as in Eq.~(\ref{eq:c1kT_tri}).
\begin{itemize}
 \item \textit{FN formula.}
 In the zero-temperature limit, the supply is $(\Ef-E)/\kb T$ below the Fermi level and zero above it.
 Integrating below the Fermi level with the linearized exponent and neglecting the contribution of the lower limit (of order $e^{-c_1\Ef}$) gives
 \begin{equation}
 \Jfn=\frac{\Ar}{\kb^2c_1^2}\,e^{-b_1}
 =\frac{\Ar(e\hbar F_\mathrm{s})^2}{8\mtun\phib\kb^2}\exp\left[-\frac{4\sqrt{2\mtun}\,\phib^{3/2}}{3e\hbar F_\mathrm{s}}\right]
 \label{eq:fn_tri}
 \end{equation}
 (supplementary data, Sect.~S3.3). This has the form of the Fowler--Nordheim formula and contains no thermal contribution.
 \item \textit{FE formula.}
 The supply is integrated without approximation. Substituting $q=e^{(E-\Ef)/\kb T}$ and extending the range to $0<q<\infty$ reduces the integral to Euler's beta integral, $\int_0^\infty q^{p-1}\ln(1+q^{-1})\,dq=\pi/(p\sin\pi p)$, and gives
 \begin{equation}
 \Jfe=\Ar T^2e^{-b_1}\,\frac{\pi}{p\sin\pi p}=\Jfn\,\frac{\pi p}{\sin\pi p},\qquad
 p\equiv c_1\kb T=\frac{\Fc}{F_\mathrm{s}}
 \label{eq:fe_tri}
 \end{equation}
 (supplementary data, Sect.~S3.4). This is Eq.~(2) of Hatakeyama et al.\cite{HatakeyamaShinohe2002}
 The factor $\pi p/\sin\pi p$ is the temperature correction due to the tail of the Fermi--Dirac distribution and tends to 1 as $T\to0$.
 The convergence condition $p<1$ corresponds to $F_\mathrm{s}>\Fc$.
 \item \textit{TFE formula.}
 Well above the Fermi level, the supply can be replaced by the Boltzmann form $e^{-(E-\Ef)/\kb T}$.
 The exponent of the integrand, $G_\mathrm{tri}(E)+(E-\Ef)/\kb T$, is minimum at the saddle point $E_\mathrm{p}$ of Eq.~(\ref{eq:Ep}). Expanding it to second order about this point and performing the two-sided Gaussian integral gives
 \begin{equation}
 \Jtfe=\frac{\Ar T}{\kb}\,\frac{e\hbar F_\mathrm{s}\sqrt\pi}{\sqrt{2\mtun\kb T}}
 \exp\left[-\frac{\phib}{\kb T}+\frac{(e\hbar F_\mathrm{s})^2}{24\mtun(\kb T)^3}\right]
 \label{eq:tfe_tri_field_equiv}
 \end{equation}
 (Appendix~A; supplementary data, Sect.~S3.5). This is Eq.~(3) of Hatakeyama et al.\cite{HatakeyamaShinohe2002} (the Hatakeyama formula); in terms of $x=E_\mathrm{A}/\kb T$ it becomes Eq.~(\ref{eq:tfe_x}).
 It requires the saddle point to lie above the Fermi level ($F_\mathrm{s}<\Fc$) with $E_\mathrm{p}-\Ef\gg\kb T$, and to lie more than about the standard deviation of the Gaussian, $\sigma_E=1/\sqrt{g''}=\sqrt{2\kb T\Delta}$, below the barrier top ($\Delta\gtrsim\sigma_E$ corresponds to $\Delta\gtrsim2\kb T$; we use $\Delta\ge3\kb T$ as a guide).
\end{itemize}

\paragraph{Quadratic barrier.}
For the quadratic barrier~(\ref{eq:Gquad}) we use $X\equiv eV_\mathrm{R}-eV_\mathrm{n}$ (so that $B=\phib+X$) and $t\equiv E_{00}/\kb T$. Energies are measured from the Fermi level ($\Ef=0$).
\begin{itemize}
 \item \textit{FE formula.}
 For $X>0$, expanding the exponent~(\ref{eq:Gquad}) to first order at the Fermi level gives
 \begin{equation}
 b_1=\frac{1}{E_{00}}\left[\sqrt{\phib B}-X\ln\frac{\sqrt B+\sqrt{\phib}}{\sqrt X}\right],\qquad
 c_1=\frac{1}{E_{00}}\operatorname{arsinh}\sqrt{\frac{\phib}{X}}
 \label{eq:B_b1c1}
 \end{equation}
 ($c_1$ is the same as in Eq.~(\ref{eq:c1_quad})), and the same integral as for the triangular barrier gives, for $p=c_1\kb T<1$,
 \begin{equation}
 \Jfe^\mathrm{quad}=\Ar T^2\,\frac{\pi}{p\sin\pi p}\,e^{-b_1}
 \label{eq:B_fe}
 \end{equation}
 (supplementary data, Sect.~S4.4).
 For $X\gg\phib$, $b_1$ and $c_1$ reduce to those of the triangular barrier, Eq.~(\ref{eq:b1c1_tri}), with correction factors of order $1+\phib/(5X)$ and $1+\phib/(3X)$, respectively.
 Equation~(\ref{eq:B_fe}) reproduces the WKB integral for the quadratic barrier within 15\% for $N_\mathrm{D}=10^{18}$--$10^{19}\,\mathrm{cm^{-3}}$ and $V_\mathrm{R}=1$--50~V (Table~S2 of the supplementary data).
 \item \textit{TFE formula.}
 With Boltzmann supply, the exponent of the integrand, $G(E)+E/\kb T$, is minimum at a saddle point $B\tanh^2t$ below the barrier top.
 Evaluating the curvature $G''$ there exactly from the exponent~(\ref{eq:Gquad}) and performing the two-sided Gaussian integral gives
 \begin{equation}
 \Jtfe^\mathrm{quad}=\frac{\Ar T}{\kb}\cdot2\sqrt{\pi E_{00}B}\,
 \mathrm{sech}\,t\sqrt{\tanh t}\;
 \exp\!\left(\frac{X}{\kb T}-\frac{B}{E_{00}}\tanh t\right)
 \label{eq:app_ps_tfe}
 \end{equation}
 (Appendix~B; supplementary data, Sects.~S4.5 and S4.6).
 For the conditions of Table~\ref{tab:app_ps}, it agrees with the total WKB current, including thermionic emission, within about 2\%. The error increases as the distance between the saddle point and the barrier top becomes comparable to the standard deviation of the Gaussian, $1/\sqrt{G''}$, and reaches about 15\% at the edge of its range of validity (Fig.~S2 of the supplementary data).
 For $t\ll1$, with $F_\mathrm{s}^2=2N_\mathrm{D}B/\epss$, it reduces to the triangular-barrier TFE formula~(\ref{eq:tfe_tri_field_equiv}) (Appendix~B).
\end{itemize}
These expressions are compared with those of PS in Sect.~\ref{sec:prev}.

\subsection{Analytical formulas and validity of the approximations}
\label{sec:approx}
\begin{figure}[htbp]
 \centering
 \includegraphics[width=0.8\linewidth]{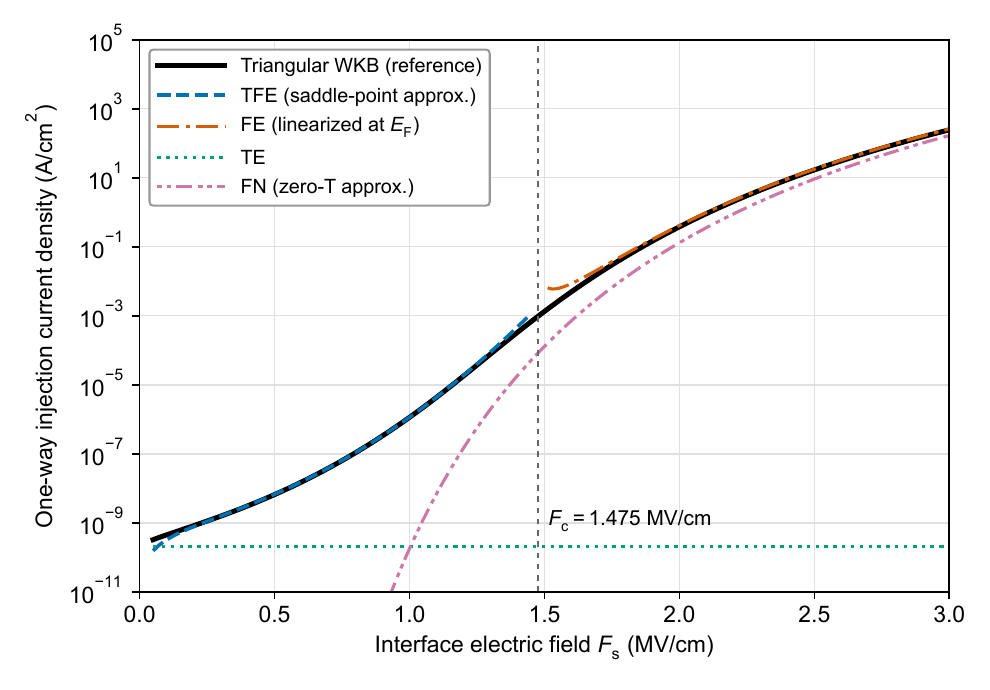}
 \caption{Comparison of the WKB numerical integral for the triangular barrier (reference, thick black line) with approximate formulas:
 TFE (saddle-point approximation, blue dashed), FE (linearization at the Fermi level, orange dash-dotted), TE (green dotted), and FN (zero-temperature approximation, purple dash-double-dotted).
 The vertical dotted line marks $\Fc=1.475\,\mathrm{MV/cm}$. $T=300\,\mathrm K$, $\phib=1.00\,\mathrm{eV}$, $\mtun=0.31m_0$.
 One-way current from the metal into the semiconductor.}
 \label{fig:approx}
\end{figure}

\paragraph{Analytical formulas.}
Figure~\ref{fig:approx} compares the WKB numerical integral for the triangular barrier, Eqs.~(\ref{eq:J}) and (\ref{eq:Gtri}),
with the closed-form TE, TFE (Eq.~(\ref{eq:tfe_tri_field_equiv})), FE (Eq.~(\ref{eq:fe_tri})), and FN (Eq.~(\ref{eq:fn_tri})) expressions.
The TFE and FE formulas agree well with the WKB integral on the $F_\mathrm{s}<\Fc$ and $F_\mathrm{s}>\Fc$ sides, respectively,
but both lose accuracy near $\Fc$ (for TFE the saddle point approaches the Fermi level; for FE the correction factor $\pi p/\sin(\pi p)$ diverges as $p\to1$).
Because the FN formula contains no thermal contribution, it strongly underestimates the current for $F_\mathrm{s}\lesssim2\,\mathrm{MV/cm}$.
Therefore, measurements that include $F_\mathrm{s}/\Fc\sim1$ must be analyzed with the WKB integral rather than with the approximate formulas.
This is consistent with the remark of Crofton and Sriram\cite{CroftonSriram1996} that the FE and TFE formulas do not reduce to one another and that neither describes the conduction accurately in the intermediate regime.

\paragraph{Triangular-barrier approximation.}
That tunneling through the quadratic barrier can be treated as tunneling through a triangular barrier follows from the asymptotic form of Weber's Eq.~(\ref{eq:weber}).
For $\mu^2\equiv\varepsilon/E_{00}\gg1$, its solutions have uniform asymptotic expansions in terms of the Airy functions $\mathrm{Ai}(w)$ and $\mathrm{Bi}(w)$ that hold over the whole range including the turning point (Eqs.~(8.11) and (11.22) of Olver\cite{Olver1959}; see also Sect.~12.10(vii) of DLMF\cite{DLMF12}).
In the forbidden region with $w\gg1$, the Airy functions vary exponentially, $\mathrm{Ai}(w)\simeq e^{-\frac23w^{3/2}}/(2\sqrt\pi\,w^{1/4})$ and $\mathrm{Bi}(w)\simeq e^{\frac23w^{3/2}}/(\sqrt\pi\,w^{1/4})$.
In the expansion, this exponent $\frac23w^{3/2}$ equals the WKB action from the turning point $z_2$ to the position $z$, $\hbar^{-1}\int_z^{z_2}\sqrt{2\mtun[U(z')-E]}\,dz'$; integrated up to the interface $z=0$ and doubled, it gives the exponent $G(E)$ of Eq.~(\ref{eq:G}) (Eq.~(\ref{eq:Gquad}) for the quadratic barrier).
Near the turning point, the variable $w$ becomes $(z_2-z)/l_\mathrm{A}(\varepsilon)$, the distance from the turning point in units of the local Airy length $l_\mathrm{A}(\varepsilon)=l_{00}(E_{00}/4\varepsilon)^{1/6}$ (supplementary data, Sect.~S6.2). This amounts to replacing the barrier near the turning point by a triangular barrier with the local slope.
The expansion parameter $\mu^{-2}=E_{00}/\varepsilon$ is below $1.5\times10^{-5}$ for the sample (the Ti/4H-SiC SBD of Ref.~\citen{Hatakeyama2003}, $N_\mathrm{D}=5\times10^{15}\,\mathrm{cm^{-3}}$).
This expansion error is distinct from the approximation of replacing the quadratic barrier by its tangent at the interface. The condition for the latter is that the barrier potential $U(z)$ can be regarded as linear (constant field) over the tunneling path from the interface to the turning point. For the quadratic barrier, the relative change of the field over this path is $z_2/W_\mathrm{d}\simeq(E_0-E)/(2B)$, so the condition is $(E_0-E)/B\ll1$ over the energy range that carries the current, i.e., the tunneling distance is much shorter than the depletion width.
With the thermal length $\lambda_\mathrm{T}=\hbar/\sqrt{\mtun\kb T}$ (about 3~nm at 300~K), the tunneling distance of the electrons carrying the TFE current is
\begin{equation}
 z_\mathrm{p}=\frac{\Delta}{eF_\mathrm{s}}=\frac{eF_\mathrm{s}\lambda_\mathrm{T}^2}{8\kb T},
 \label{eq:zp}
\end{equation}
which increases in proportion to the field and reaches its maximum $\phib/(e\Fc)$ at $F_\mathrm{s}=\Fc$.
For the sample (30\degC), $z_\mathrm{p}\approx1.3$--6~nm (50--950~V), much shorter than the depletion width (3--10~$\mu$m), so this condition is well satisfied.
In heavily doped layers, by contrast, the tunneling distance becomes comparable to the depletion width, and the curvature correction of Eq.~(\ref{eq:c1_quad}) is needed.

\paragraph{WKB approximation.}
For a triangular barrier with a uniform field, the solutions of the Schr\"odinger equation are expressed by Airy functions, and the transmission probability can be obtained exactly\cite{Gundlach1966} (supplementary data, Sect.~S6.3).
In the exact solution, the exponential factor $e^{-G}$ is multiplied by a factor that arises from matching the wave functions at the metal--semiconductor interface (hereafter, the interface-matching factor). The ratio of the exact transmission probability to that of the WKB approximation including the same interface-matching factor is essentially a function of $\xi_0=(E_0-E)/E_\mathrm{A}$ alone; varying $E_\mathrm{A}$ over 0.05--0.2~eV changes it by less than 0.06.
The ratio deviates from unity by less than 10\% for $\xi_0\ge2$.
For the field and temperature conditions of the sample (0.3--1.4~MV/cm, 30--160\degC; Table~S5 of the supplementary data), the Airy and WKB currents differ by at most 0.05~dex when the WKB calculation includes the interface-matching factor below the barrier top and the step transmission above it (dex denotes the unit of $\log_{10}$; 1~dex corresponds to a factor of 10).
The exponential factor $e^{-G}$ of the WKB approximation is therefore accurate in this range.
The current fits in Sect.~\ref{sec:exp}, in contrast, use $e^{-G}$ below the barrier top and unit transmission above it, without the interface-matching factor, as is usual in TE and TFE theory (it is not treated, as with the correction to the Richardson constant mentioned in Sect.~\ref{sec:model}).
In the following, the difference between calculation and measurement is expressed by the residual $r=\log_{10}(J_\mathrm{calc}/J_\mathrm{meas})$, and its magnitude by the root-mean-square error, $\mathrm{RMSE}=\sqrt{\langle r^2\rangle}$.

\subsection{Re-examination of previously reported closed-form expressions and transition conditions}
\label{sec:prev}
\paragraph{Formulas for the triangular barrier.}
The FE and TFE formulas of Hatakeyama et al.\cite{HatakeyamaShinohe2002} are the triangular-barrier expressions~(\ref{eq:fe_tri}) and (\ref{eq:tfe_tri_field_equiv}) of Sect.~\ref{sec:closed}.
The FE formula is algebraically identical to Eqs.~(37)--(39) of PS\cite{PadovaniStratton1966}, which approximate their reverse FE formula in the limit where the reverse voltage greatly exceeds the barrier height (supplementary data, Sect.~S4.4).
The TFE formula agrees with the $E_{00}\ll\kb T$ limit of Eq.~(\ref{eq:app_ps_tfe}), the TFE current of the quadratic barrier evaluated correctly by the saddle-point method; under the conditions of low-doped layers the two differ by less than 0.3\% (Appendix~B).

\paragraph{Formulas for the quadratic barrier.}
PS\cite{PadovaniStratton1966} gave reverse FE and TFE formulas for the quadratic barrier.
Table~\ref{tab:closed} compares their formulas with those re-examined in this work (derivations in Appendix~B and supplementary data, Sect.~S4).
Their general FE formula (Eq.~(3) of PS with their Eqs.~(34) and (35)) agrees with our Eq.~(\ref{eq:B_fe}) for $eV_\mathrm{n}=0$.
Equations~(37)--(39) of PS, which they used to analyze their experiments, are, however, the leading terms in the limit where the reverse voltage greatly exceeds the barrier height ($\phib/X\ll1$), and are identical to the triangular-barrier FE formula.
Since electrons near the Fermi level tunnel at a depth $\phib$ below the barrier top, $\phib/X\ll1$ corresponds to the condition $E_0-E\ll B$ of the triangular-barrier approximation in Sect.~\ref{sec:approx}.
FE, however, dominates in heavily doped layers with large $E_{00}$, where high interface fields are reached at reverse voltages of a few volts; $\phib/X$ is then not small, and the curvature of the barrier cannot be neglected.
Under such conditions PS effectively adopt the triangular-barrier approximation; for example, for $N_\mathrm{D}=10^{19}\,\mathrm{cm^{-3}}$ and $V_\mathrm{R}=1$--10~V ($\phib/X=1.0$--0.1), Eq.~(39) of PS overestimates the WKB current of the quadratic barrier 1.4--7.9-fold.
In their TFE formula, the exponent (Eqs.~(45) and (47)--(49) of PS) agrees, but, for the low-doped cases at 300--433~K in Table~\ref{tab:app_ps}, Eq.~(50) of PS gives 2.5--3.6 times the total WKB current, because the curvature at the saddle point is evaluated with an approximation valid only for a saddle point well below the barrier top and only one side of the Gaussian integral is taken.
\begin{table}[htbp]
 \caption{\label{tab:closed}Reverse FE and TFE formulas for the quadratic barrier: formulas of PS\cite{PadovaniStratton1966} and of this work. $X=eV_\mathrm{R}-eV_\mathrm{n}$, $B=\phib+X$, $t=E_{00}/\kb T$. The TFE prefactors are given without the common factor $\Ar T/\kb$. Errors are relative to the WKB numerical integral, Eq.~(\ref{eq:J}), for $\phib=1.0\,\mathrm{eV}$, $\mtun=0.31m_0$, and $\varepsilon_\mathrm{r}=10.03$ (Table~\ref{tab:app_ps}; supplementary data, Table~S2).}
 \renewcommand{\arraystretch}{1.6}
 \begin{indented}
 \item[]\footnotesize\begin{tabular}{@{}>{\raggedright\arraybackslash}p{0.11\linewidth}>{\raggedright\arraybackslash}p{0.30\linewidth}>{\raggedright\arraybackslash}p{0.29\linewidth}>{\raggedright\arraybackslash}p{0.19\linewidth}@{}}
 \br
 & PS & This work & Difference; error relative to WKB\\
 \mr
 FE: $b_1$ & $\dfrac{2\phib^{3/2}}{3E_{00}\sqrt{\phib+eV_\mathrm{R}}}$ (PS Eq.~(37)) & $\dfrac{1}{E_{00}}\Bigl[\sqrt{\phib B}$\newline\hspace*{1.2em}$-X\ln\dfrac{\sqrt B+\sqrt{\phib}}{\sqrt X}\Bigr]$ (Eq.~(\ref{eq:B_b1c1})) & PS: leading term for $\phib/X\ll1$; $eV_\mathrm{n}$ omitted\\
 FE: $c_1$ & $\dfrac{1}{E_{00}}\sqrt{\dfrac{\phib}{\phib+eV_\mathrm{R}}}$ (PS Eq.~(38)) & $\dfrac{1}{E_{00}}\operatorname{arsinh}\sqrt{\dfrac{\phib}{X}}$ (Eq.~(\ref{eq:B_b1c1})) & Same as above\\
 FE: current & PS Eq.~(39) & Eq.~(\ref{eq:B_fe}) & $N_\mathrm{D}=10^{18}$--$10^{19}\,\mathrm{cm^{-3}}$: PS $\times1.14$--7.9; this work $\times1.05$--1.14\\
 TFE: exponent & $\dfrac{X}{\kb T}-\dfrac{B}{E_{00}}\tanh t$ (PS Eqs.~(45)--(49)) & Same & Agree\\
 TFE: prefactor & $\sqrt{\pi E_{00}(eV_\mathrm{R}+\phib\,\mathrm{sech}^2t)}$ (PS Eq.~(50)) & $2\sqrt{\pi E_{00}B}\,\mathrm{sech}\,t\sqrt{\tanh t}$ (Eq.~(\ref{eq:app_ps_tfe})) & Curvature approximation and one-sided Gaussian integral; PS $-14$\% to $+261$\%, this work $+0.5$ to $+2.0$\% (conditions of Table~\ref{tab:app_ps}, total current)\\
 \br
 \end{tabular}
 \end{indented}
\end{table}

\paragraph{Emission regimes and transition fields.}
The TFE/FE boundary that Hatakeyama et al.\cite{HatakeyamaShinohe2002} mapped numerically in the plane of interface field and barrier height (300~K, $\mtun=0.45m_0$; about 1.25~MV/cm at a barrier height of 0.5~eV and
about 2.2~MV/cm at 1.5~eV) agrees with $\Fc$ for the same mass (1.26 and 2.18~MV/cm). The independence of their TE/TFE boundary at about 0.3~MV/cm from the barrier height is also explained by Eq.~(\ref{eq:tfe_x}).
Li et al.\cite{Li2020} found for $\beta$-Ga$_2$O$_3$ ($\mtun=0.31m_0$) that the field at which the TE and tunneling currents are equal depends little on the barrier height and doping, and called it a universal transition field.
This corresponds to $J_\mathrm{tun}=\Jte$ ($f(x)=1$) in Table~\ref{tab:regime}.
Their value (0.27~MV/cm at 300~K, from their fitted expression) is about three times our 0.085~MV/cm without the image force; the ratio is about 2.8--3.4 between 100 and 800~K (3.36, 3.14, and 2.81 at 100, 300, and 800~K).
Li et al.\ include the image force in the barrier and separate the TE and tunneling currents at the top of the lowered barrier. The image force therefore not only increases the TE current but also changes the shape of the tunneling barrier and the range of the tunneling integral.
Recalculating the equal-current field with the image-force exponent~(\ref{eq:GBL}) ($m_\mathrm{tun}=0.31m_0$, $\varepsilon_\mathrm{r}=10.03$, Boltzmann supply, separation at the lowered barrier top), we obtain 0.049, 0.220, and 0.85~MV/cm at 100, 300, and 800~K, which are 0.82--0.89 times their values.
The image force thus explains most of the difference.

\paragraph{TFE/FE transition in heavily doped layers.}
Hara et al.\cite{Hara2022} analyzed the reverse characteristics of Ni/heavily doped 4H-SiC SBDs by WKB calculations for a quadratic barrier including the image force,
and defined the TFE/FE transition as the point at which the peak of the current distribution, $E_\mathrm{peak}$, falls to 26~meV above the metal Fermi level $E_\mathrm{Fm}$.
They reported that the interface field at the transition, $F_\mathrm{max}$, depends on $N_\mathrm{D}$ (2.9~MV/cm for $1.2\times10^{19}\,\mathrm{cm^{-3}}$ and 2.1~MV/cm for $2.6\times10^{18}\,\mathrm{cm^{-3}}$),
whereas the local field at $E_\mathrm{peak}$, $F_\mathrm{peak}$, is about 1.6~MV/cm, nearly independent of $N_\mathrm{D}$.
Table~\ref{tab:hara} shows our recalculation of the transition points under their conditions. It reproduces the reported values, but when $N_\mathrm{D}$ is extended to $2\times10^{17}$--$5\times10^{19}\,\mathrm{cm^{-3}}$,
$F_\mathrm{max}$ varies from 2.1 to 5.1~MV/cm and $F_\mathrm{peak}$ from 2.0 to 1.2~MV/cm.
In contrast, $c_1\kb T$ at the transition is nearly constant at 0.87--0.88 for $N_\mathrm{D}\le1.2\times10^{19}\,\mathrm{cm^{-3}}$ and stays within 0.86--0.93 with or without the image force and for a different barrier height.
This value agrees with $c_1\kb T=0.86$, derived from the product of supply and transmission as the condition for the peak to lie at $\Ef+\kb T$ (supplementary data, Sect.~S6.6).
At $5\times10^{19}\,\mathrm{cm^{-3}}$, where the transition occurs at nearly zero bias and the back flow from the semiconductor is not negligible, $c_1\kb T$ is 0.80 (0.69--0.81 for all conditions; supplementary data, Table~S6).
The transition condition of Hara et al.\ is therefore described in a unified way by $c_1\kb T\approx1$, the extension of $F_\mathrm{s}=\Fc$ to general barrier shapes.
The interpretation of the nearly constant $F_\mathrm{peak}$ is given in supplementary data, Sect.~S6.6, and that of their constant mass $0.30m_0$ in Sect.~\ref{sec:meff_interp}.

\begin{table}[htbp]
 \caption{\label{tab:hara}Recalculation of the TFE/FE transition condition of Hara et al.\cite{Hara2022} ($E_\mathrm{peak}=E_\mathrm{Fm}+26\,\mathrm{meV}$).
 Quadratic barrier with image force, WKB, $T=300\,\mathrm K$, $\phib=1.45\,\mathrm{eV}$, $\mtun=0.30m_0$, $\epss=10.32\varepsilon_0$.
 $F_\mathrm{max}$: interface field; $F_\mathrm{peak}$: field at the outer turning point at $E_\mathrm{peak}$; $c_1\kb T$: Eq.~(\ref{eq:c1}) evaluated numerically at $E_\mathrm{Fm}$. Values reported by Hara et al.\ are in parentheses. $^\mathrm{a}$Bulk $E_\mathrm{C}-E_\mathrm{F}$ from Fermi--Dirac statistics ($N_\mathrm{D}>N_\mathrm{C}$); the other rows use the non-degenerate expression, which differs from the Fermi--Dirac value by at most 0.006~eV (at $1.2\times10^{19}\,\mathrm{cm^{-3}}$). The transition lies within about 1~mV of zero bias, and its sign (forward or reverse) cannot be determined at the resolution of this calculation; at exactly zero bias the two one-way currents cancel and the peak of the net current distribution is not defined. With the non-degenerate expression, this row would give $V_\mathrm{R}=0.01$~V, 5.11, 1.13, and 0.85.}
 \begin{indented}
 \item[]\begin{tabular}{@{}rrrrr}
 \br
 $N_\mathrm{D}$ (cm$^{-3}$) & $V_\mathrm{R}$ (V) & $F_\mathrm{max}$ (MV/cm) & $F_\mathrm{peak}$ (MV/cm) & $c_1\kb T$\\
 \mr
 $2.0\times10^{17}$ & 59 & 2.06 & 2.03 & 0.88\\
 $1.0\times10^{18}$ & 11 & 2.09 & 1.97 & 0.87\\
 $2.6\times10^{18}$ & 4.1 & 2.24 (2.1) & 1.93 & 0.87\\
 $1.2\times10^{19}$ & 0.59 & 2.93 (2.9) & 1.61 (1.6) & 0.88\\
 $5.0\times10^{19}$\,$^\mathrm{a}$ & $\approx0$ & 5.14 & 1.24 & 0.80\\
 \br
 \end{tabular}
 \end{indented}
\end{table}

\section{Effects of barrier shape and tunneling effective mass}
\label{sec:bl}
\subsection{Change of barrier shape by the image force}

Including the image force (hereafter BL, for barrier lowering)\cite{SzeNg2007}, $U_\mathrm{BL}(z)=E_0-eF_\mathrm{s}z-e^2/(16\pi\epss z)$,
and the barrier top is lowered by $\Delta U=\sqrt{e^3F_\mathrm{s}/(4\pi\epss)}$. The action integral can be written as
\begin{equation}
 G_\mathrm{BL}(E)=v(y)\,G_\mathrm{tri}(E),\qquad
 v(y)=\sqrt{1+y}\left[\mathsf{E}(\nu)-y\mathsf{K}(\nu)\right],
 \label{eq:GBL}
\end{equation}
with $y=\Delta U/(E_0-E)$ and $\nu=(1-y)/(1+y)$\cite{MurphyGood1956,Forbes2006} (supplementary data, Sect.~S5); $\mathsf{K}$ and $\mathsf{E}$ are complete elliptic integrals with parameter $\nu$.
In this case the thermionic component is also evaluated with $\phib-\Delta U$ as the barrier.

\begin{figure}[htbp]
 \centering
 \includegraphics[width=0.8\linewidth]{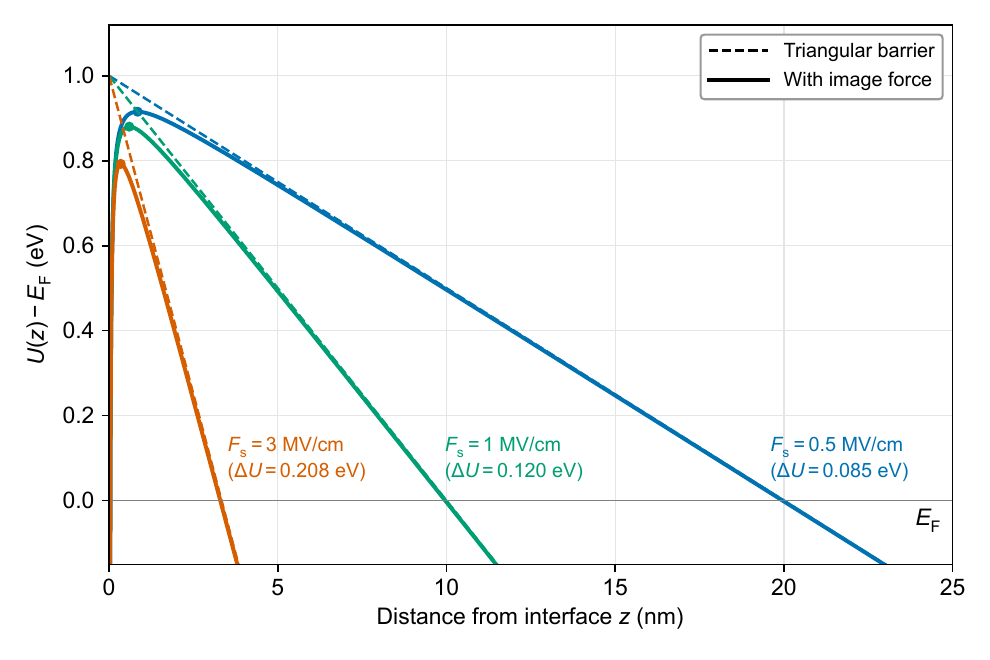}
 \caption{Triangular barrier (dashed) and barrier including the image force (solid) for interface fields of 0.5 (blue), 1 (green), and 3~MV/cm (orange). Circles mark the top of the image-force barrier.
 $\phib=1.00\,\mathrm{eV}$, $\varepsilon_\mathrm{r}=10.03$. The divergence of the image potential as $z\to0$ indicates the limit of the continuum approximation.}
 \label{fig:bl_shape}
\end{figure}

Figure~\ref{fig:bl_shape} compares the triangular barrier with the barrier including the image force.
The image force not only lowers the barrier top by $\Delta U$ but also creates a turning point on the interface side, narrowing the forbidden region.
What matters is not the geometrical area of the barrier but the reduction of the action integral~(\ref{eq:G}), which is expressed by $v(y)<1$.
Lowering only the barrier height for TE by $\Delta U$ (hereafter simple BL) does not include this change of shape.

\begin{figure}[htbp]
 \centering
 \includegraphics[width=0.8\linewidth]{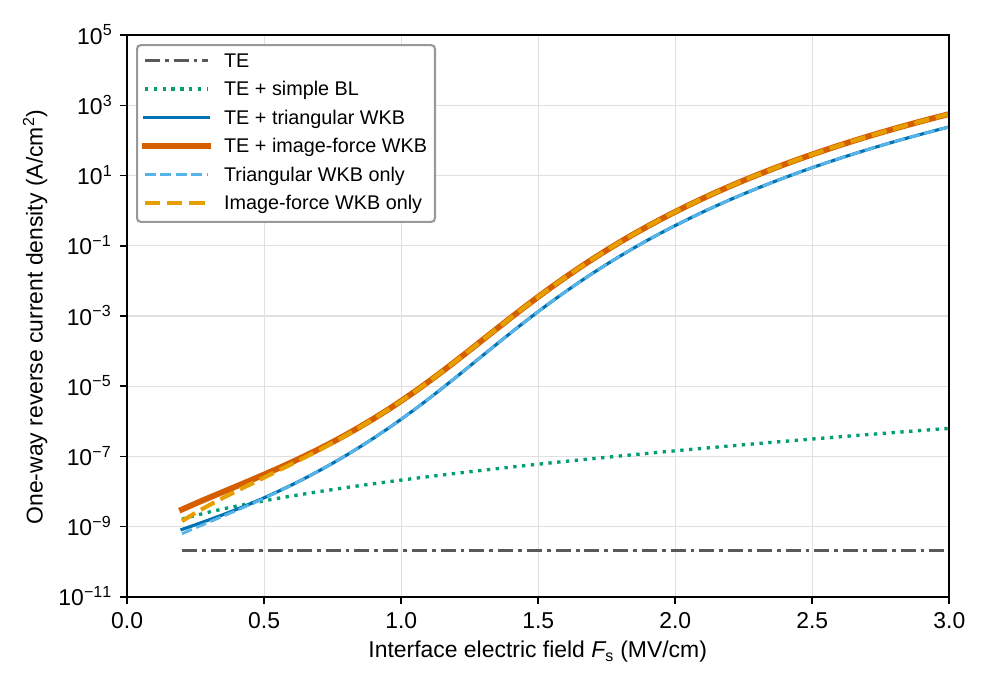}
 \caption{Interface-field dependence of the one-way reverse current density calculated with TE, simple BL, triangular-barrier WKB, and WKB including the image force (see legend; dashed curves show the tunneling components only).
 $T=300\,\mathrm K$, $\phib=1.00\,\mathrm{eV}$, $\varepsilon_\mathrm{r}=10.03$, $\mtun=0.31m_0$.}
 \label{fig:bl_current}
\end{figure}

Figure~\ref{fig:bl_current} and Table~\ref{tab:bl} show the increase of the current due to the image force.
At low fields, the current above the lowered barrier top is counted as the TE component, so the ratio of the tunneling components alone does not represent the actual increase.
Comparing total currents including TE, the enhancement at 300~K is largest (a factor of 4.6) near 0.5~MV/cm
and decreases with increasing field at higher fields (2.7 at 1.4~MV/cm and 2.3 at 3~MV/cm).
The decrease at high fields occurs because the triangular barrier itself becomes thin, so the relative reduction of the action integral by the image force becomes smaller.
The enhancement also depends on temperature: at 433~K its maximum moves to about 0.7--1~MV/cm, and its variation over 0.3--1.4~MV/cm is small (3.0--3.8).
In the field range of the sample (0.3--1.4~MV/cm), BL weakens the voltage dependence more at lower temperatures and also changes the temperature dependence through the temperature variation of the enhancement.
These effects explain why adding BL to a constant-mass model moves the calculation further away from the measurements in Sect.~\ref{sec:exp}.

\begin{table}[htbp]
 \caption{\label{tab:bl}Current enhancement by the image force, $J_\mathrm{BL}/J_\mathrm{tri}$, for the tunneling components only and for the total current including TE. $\phib=1.00\,\mathrm{eV}$, $\varepsilon_\mathrm{r}=10.03$, $\mtun=0.31m_0$.}
 \begin{indented}
 \item[]\begin{tabular}{@{}rrccc}
 \br
 & & Tunneling only & \multicolumn{2}{c}{Total current}\\
 $F_\mathrm{s}$ (MV/cm) & $\Delta U$ (eV) & 300~K & 300~K & 433~K\\
 \mr
 0.2 & 0.054 & 2.30 & 3.70 & 2.60\\
 0.3 & 0.066 & 3.02 & 4.29 & 3.01\\
 0.5 & 0.085 & 3.90 & 4.60 & 3.55\\
 0.7 & 0.100 & 3.94 & 4.17 & 3.77\\
 1.0 & 0.120 & 3.30 & 3.32 & 3.66\\
 1.4 & 0.142 & 2.70 & 2.70 & 3.17\\
 2.0 & 0.169 & 2.46 & 2.46 & 2.62\\
 3.0 & 0.208 & 2.32 & 2.32 & 2.36\\
 \br
 \end{tabular}
 \end{indented}
\end{table}

\subsection{Energy-dependent tunneling effective mass}

The wave number of an electron in the band gap is imaginary (in general complex), and its dispersion is given by the complex band structure\cite{Heine1963,Chang1982}. The imaginary wave number deviates from the extrapolation of the parabolic band. Evaluating the WKB attenuation with the imaginary branch of the band structure, and obtaining this branch by continuing a multiband $\mathbf{k}\cdot\mathbf{p}$ model to complex $k$, goes back to Kane\cite{Kane1961}.
From the complex band structure of 4H-SiC, Murakami\cite{Murakami2023} gave, for the energy $\mathcal{E}=E-U(z)$ measured from the local conduction-band edge (in eV, $-2\leq\mathcal{E}\leq0$),
\begin{equation}
 \frac{\mtun(\mathcal{E})}{m_0}=0.0171\mathcal{E}^3+0.0798\mathcal{E}^2+0.1772\mathcal{E}+0.3241.
 \label{eq:mE}
\end{equation}
$\mtun$ decreases with depth: $0.32m_0$ at the conduction-band edge, $0.25m_0$ at 0.5~eV below it, and $0.21m_0$ at 1~eV below it.
When this mass is used, the integrand of Eq.~(\ref{eq:G}) is evaluated numerically with $\mtun\to\mtun(\mathcal{E})$; for $\mathcal{E}<-2$~eV, outside the range of Eq.~(\ref{eq:mE}), the value at $-2$~eV is used. We refer to this energy-dependent mass as $\mtun(E)$ when naming models.

\subsection{Consequences for the field and temperature dependence}
\label{sec:consequence}
\begin{figure}[htbp]
 \centering
 \includegraphics[width=0.65\linewidth]{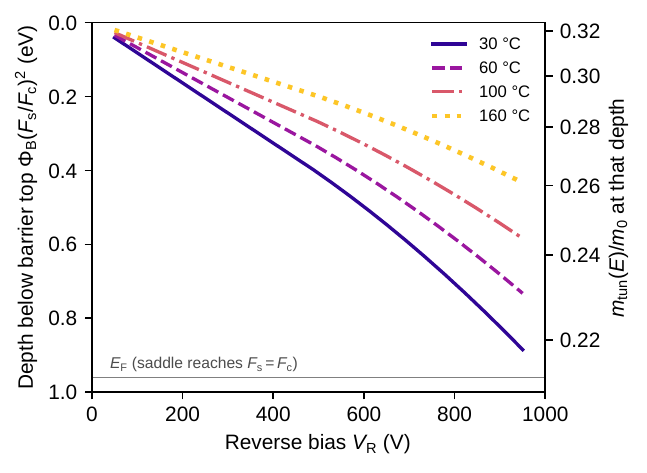}
 \caption{Depth below the barrier top of the energy of the current-carrying electrons, $\phib(F_\mathrm{s}/\Fc)^2$ (left axis; constant-mass reference model without image force), and $\mtun(E)$ at that depth (right axis; Eq.~(\ref{eq:mE})), for the sample whose measured characteristics are analyzed in Sect.~\ref{sec:exp}.
 The interface field including punch-through, Eq.~(\ref{eq:Fs}), was used. Line colors follow the temperature scale of Fig.~\ref{fig:fit}.}
 \label{fig:depth}
\end{figure}

In the constant-mass reference model without image force, Eq.~(\ref{eq:Ep}), the current-carrying electrons tunnel at $\Delta=\phib(F_\mathrm{s}/\Fc)^2$ below the barrier top.
Taking as an example the sample whose measured characteristics are analyzed in Sect.~\ref{sec:exp} ($N_\mathrm{D}=5\times10^{15}\,\mathrm{cm^{-3}}$, 50--950~V, 30--160\degC; thick bar in Fig.~\ref{fig:regime}),
Fig.~\ref{fig:depth} and Table~\ref{tab:depth} show $F_\mathrm{s}/\Fc$, $\Delta$, and $\mtun(E)$ at that depth.
As the voltage increases, the tunneling energy moves deeper into the gap and the effective mass decreases from $0.32m_0$ to $0.22m_0$.
Because a smaller mass gives a larger transmission probability, $\mtun(E)$ increases the current more at higher voltages and strengthens the voltage dependence.
This depth-dependent change of mass is not contained in constant-mass models.
Moreover, because the depth is larger at lower temperatures ($\Fc\propto T$), the effect of $\mtun(E)$ is stronger at lower temperatures.

\begin{table}[htbp]
 \caption{\label{tab:depth}$F_\mathrm{s}/\Fc$, tunneling depth $\Delta=\phib(F_\mathrm{s}/\Fc)^2$, and $\mtun(E)$ at that depth for the sample of Sect.~\ref{sec:exp} ($\phib=0.96\,\mathrm{eV}$). The depth is an estimate from the reference model with a constant mass of $0.31m_0$ and no image force; it is not the equivalent mass along the whole path.}
 \begin{indented}
 \item[]\begin{tabular}{@{}rrrrrr}
 \br
 $T$ ($^\circ$C) & $V_\mathrm{R}$ (V) & $F_\mathrm{s}$ (MV/cm) & $F_\mathrm{s}/\Fc$ & $\Delta$ (eV) & $\mtun(E)/m_0$\\
 \mr
 30--160 & 50 & 0.30 & 0.14--0.21 & 0.02--0.04 & 0.32\\
 160 & 950 & 1.40 & 0.67 & 0.43 & 0.26\\
 100 & 950 & 1.40 & 0.78 & 0.58 & 0.24\\
 60 & 950 & 1.40 & 0.87 & 0.73 & 0.23\\
 30 & 950 & 1.40 & 0.96 & 0.88 & 0.22\\
 \br
 \end{tabular}
 \end{indented}
\end{table}

The saddle depth used here is not obtained self-consistently for the model including Eq.~(\ref{eq:mE}); it is a reference quantity for understanding the effect.
Currents including the energy-dependent mass and the image force are calculated from the numerical integral~(\ref{eq:J}), not by substituting approximate saddle-point formulas.
For a constant mass, the enhancement by the image force depends on field and temperature, as shown in Table~\ref{tab:bl}, whereas the energy-dependent mass becomes important on the high-field, low-temperature side where deep energies are involved.
Below we compare how these different effects appear in the measured voltage and temperature dependence.

\section{Application to measured reverse characteristics}
\label{sec:exp}

\subsection{Sample and measurements}
We used the reverse current--voltage characteristics of the Ti/4H-SiC SBD of Ref.~\citen{Hatakeyama2003}.
The SBDs were fabricated on n-type 4H-SiC epitaxial wafers supplied by Cree Research.
The drift layer has a donor concentration of about $5\times10^{15}\,\mathrm{cm^{-3}}$ and a thickness of 10~$\mu$m;
the Schottky electrode is 300~$\mu$m in diameter, the edge termination is a junction termination extension (JTE) formed by Al implantation, the breakdown voltage is about 1200~V, and the specific on-resistance is about 2.5~m$\Omega$\,cm$^2$.
At room temperature, the reverse characteristics of different devices coincide well (except for devices showing a sudden increase due to accidental crystal defects), indicating that the leakage originates from a mechanism intrinsic to the Schottky interface\cite{Hatakeyama2003}.
The barrier height obtained from the forward characteristics is 1.0~eV; the previous reverse analysis, which fitted $\mtun$, $\phib$, and $A^*$ simultaneously with a constant-mass TFE formula, gave $\mtun=0.25m_0$, $\phib=0.96\,\mathrm{eV}$, and $A^*=150\,\mathrm{A\,cm^{-2}K^{-2}}$\cite{Hatakeyama2003}.
The reverse current--voltage characteristics were measured with a DC source/monitor (Agilent 4142B), using the same measurement system as in Ref.~\citen{Hatakeyama2004}.
The measurement temperatures were 30, 40, 60, 80, 100, 120, 140, and 160\degC, and the voltage range was 0--1200~V.
Only some of the temperatures were reported previously; here we reanalyze the full data set including the unpublished temperatures.
Above 1100~V the current saturates at about 1.6~A/cm$^2$ because of the current limit of the instrument.

\subsection{Analysis method}
The interface field of Eq.~(\ref{eq:Fs}) was used in Eq.~(\ref{eq:J}), and the current was calculated with the following four models:
\begin{itemize}
 \item[(a)] WKB: triangular barrier~(\ref{eq:Gtri}), $\mtun=0.31m_0$;
 \item[(b)] WKB+BL: barrier including the image force~(\ref{eq:GBL}), $\mtun=0.31m_0$;
 \item[(c)] WKB+$\mtun(E)$: triangular barrier, Eq.~(\ref{eq:mE});
 \item[(d)] WKB+BL+$\mtun(E)$: barrier including the image force, Eq.~(\ref{eq:mE}).
\end{itemize}
All models include the TE component. The triangular barrier is used because, in the drift layer of the sample ($B\gtrsim50\,\mathrm{eV}$), $(E_0-E)/B\ll1$ holds over the energy range that carries the current, so that the difference from the quadratic barrier is negligible (Sect.~\ref{sec:index}).
The only fitting parameter is the barrier height $\phib$, common to all temperatures.
The evaluation used 50--950~V at 25~V intervals (37 points per temperature, 296 points in total), and the logarithmic median within $\pm12.5\,\mathrm V$ of each point was taken to suppress noise.
The figure of merit is the RMSE of the residuals $r$ of $\log_{10}J$ defined in Sect.~\ref{sec:approx}, in dex.
An RMSE of 0.1~dex corresponds to a typical current ratio of $10^{0.1}=1.26$.
The region above 950~V was excluded because of the excess current discussed below and the current limit.
The calculation of the interface field and the optimization procedure are given in Appendix~C, and the sensitivity to the sample parameters in Sect.~\ref{sec:sensitivity}.

\subsection{Current characteristics and residuals}
\begin{figure}[htbp]
 \centering
 \includegraphics[width=\linewidth]{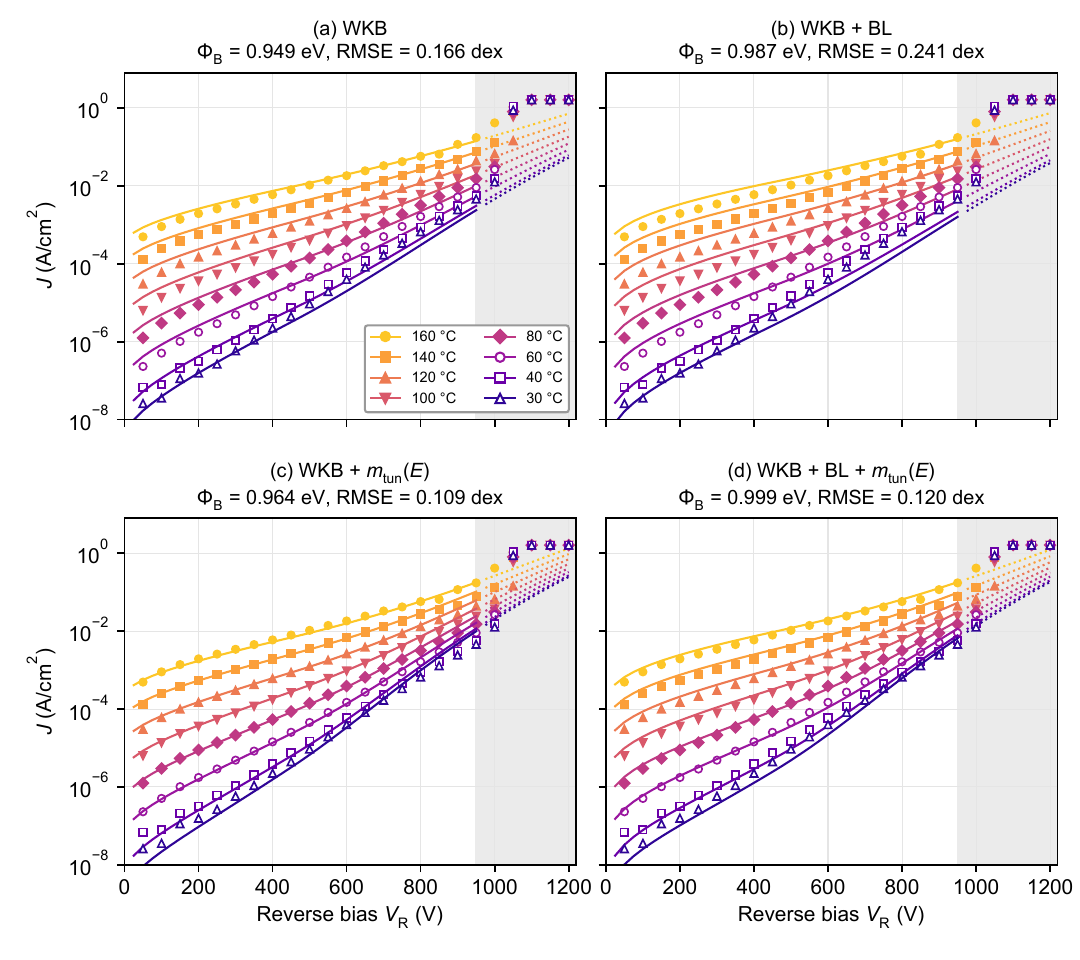}
 \caption{Measured reverse current of the Ti/4H-SiC SBD (symbols, 30--160\degC) and calculations with the four models (solid lines). Symbol and line colors indicate temperature, from purple (low) to yellow (high), common to all panels.
 $N_\mathrm{D}=5\times10^{15}\,\mathrm{cm^{-3}}$. Gray symbols lie outside the fitting range (above 950~V), and dotted lines are extrapolations of the calculations.
 The saturation above 1100~V is due to the current limit.}
 \label{fig:fit}
\end{figure}

\begin{figure}[htbp]
 \centering
 \includegraphics[width=\linewidth]{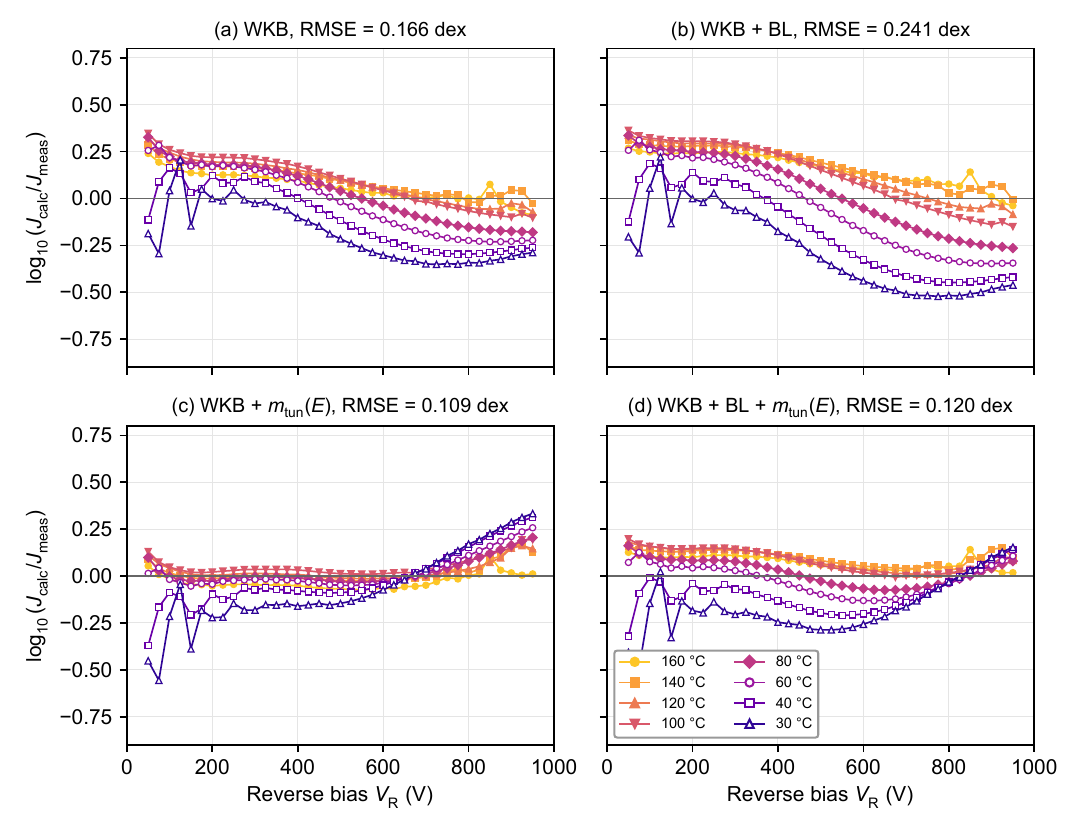}
 \caption{Voltage dependence of the residuals $\log_{10}(J_\mathrm{calc}/J_\mathrm{meas})$ of the four models ($N_\mathrm{D}=5\times10^{15}\,\mathrm{cm^{-3}}$).
 Colors and symbols are as in Fig.~\ref{fig:fit}. The scatter on the low-voltage side at 30 and 40\degC\ is due to noise near the measurement floor.}
 \label{fig:resid}
\end{figure}

Table~\ref{tab:fit} lists the fitting results of the four models, Fig.~\ref{fig:fit} compares measured and calculated currents, and Fig.~\ref{fig:resid} shows the residuals.
Considering the uncertainty in the donor concentration, results with $N_\mathrm{D}$ changed by $\pm10\%$ are also listed.

\begin{table}[htbp]
 \caption{\label{tab:fit}Fitting results of the four models: barrier height $\phib$ (eV) / RMSE (dex).}
 \begin{indented}
 \item[]\begin{tabular}{@{}lccc}
 \br
 Model & $N_\mathrm{D}=4.5\times10^{15}$ & $5.0\times10^{15}$ & $5.5\times10^{15}\,\mathrm{cm^{-3}}$\\
 \mr
 (a) WKB & 0.939 / 0.218 & 0.949 / 0.166 & 0.959 / 0.118\\
 (b) WKB+BL & 0.978 / 0.289 & 0.987 / 0.241 & 0.997 / 0.193\\
 (c) WKB+$\mtun(E)$ & 0.952 / \textbf{0.101} & 0.964 / \textbf{0.109} & 0.976 / 0.159\\
 (d) WKB+BL+$\mtun(E)$ & 0.988 / 0.162 & 0.999 / 0.120 & 1.011 / \textbf{0.114}\\
 \br
 \end{tabular}
 \end{indented}
\end{table}

\begin{table}[htbp]
 \caption{\label{tab:perT}RMSE (dex) for each temperature and voltage change of the residual $\Delta r_V$ (dex; linear fit scaled to 900~V). $N_\mathrm{D}=5\times10^{15}\,\mathrm{cm^{-3}}$.}
 \begin{indented}
 \item[]\begin{tabular}{@{}rcccccc}
 \br
 & \multicolumn{4}{c}{RMSE} & \multicolumn{2}{c}{$\Delta r_V$}\\
 $T$ ($^\circ$C) & (a) & (b) & (c) & (d) & (c) & (d)\\
 \mr
 160 & 0.097 & 0.181 & 0.050 & 0.080 & $+0.04$ & $-0.09$\\
 140 & 0.126 & 0.202 & 0.054 & 0.106 & $+0.08$ & $-0.08$\\
 120 & 0.143 & 0.204 & 0.055 & 0.102 & $+0.06$ & $-0.15$\\
 100 & 0.161 & 0.211 & 0.073 & 0.102 & $+0.08$ & $-0.16$\\
 80 & 0.154 & 0.202 & 0.070 & 0.069 & $+0.14$ & $-0.14$\\
 60 & 0.174 & 0.236 & 0.090 & 0.079 & $+0.21$ & $-0.10$\\
 40 & 0.193 & 0.288 & 0.149 & 0.136 & $+0.45$ & $+0.13$\\
 30 & 0.241 & 0.355 & 0.213 & 0.218 & $+0.62$ & $+0.31$\\
 \br
 \end{tabular}
 \end{indented}
\end{table}

The voltage change of the residual $r$ (Sect.~\ref{sec:approx}) in Table~\ref{tab:perT}, $\Delta r_V$, is obtained by a least-squares linear fit $r\simeq aV_\mathrm{R}+b$ of the residuals over 50--950~V, with $\Delta r_V=a\times900\,\mathrm V$.
It represents the systematic voltage dependence over the whole evaluation range rather than the difference between the residuals at the end points.
``Residual change'' below refers to this quantity.

The results can be summarized as follows.
\begin{enumerate}
 \item \textbf{With nominal parameters, $\mtun(E)$ improves the reproducibility without a mass-adjustment parameter.}
 For the nominal $N_\mathrm{D}=5.0\times10^{15}\,\mathrm{cm^{-3}}$ and for $4.5\times10^{15}\,\mathrm{cm^{-3}}$, the RMSEs of (c) and (d) are smaller than those of (a) and (b).
 With the constant mass of (a) and (b), the residual change between 50 and 950~V is $-0.4$ to $-0.8$~dex at 60--100\degC,
 i.e., the calculation underestimates the voltage dependence (Figs.~\ref{fig:resid}(a) and \ref{fig:resid}(b)). $\mtun(E)$ reduces this systematic residual change.
 With $N_\mathrm{D}$ increased by 10\%, the RMSE of (a) also decreases to 0.118~dex, comparable to (d) (0.114~dex).
 Increasing $N_\mathrm{D}$ raises the interface field, and introducing $\mtun(E)$ raises the tunneling probability at deep energies; because both strengthen the voltage dependence of the current, they are correlated in the fit.
 However, the slopes of the linear fits to the residuals of (a) and (b) remain negative at all temperatures even when $N_\mathrm{D}$ is changed by $\pm10\%$,
 so in this comparison with the mass fixed at $0.31m_0$ the voltage dependence remains insufficient.
 \item \textbf{Adding BL with the mass fixed at $0.31m_0$ worsens the agreement.}
 Model (b) has the largest RMSE for every $N_\mathrm{D}$. Its residuals have a more negative voltage dependence than those of (a) at all temperatures (e.g., $-0.63$~dex at 30\degC\ compared with $-0.40$~dex for (a)),
 and the spread of the mean deviation among temperatures widens from 0.29~dex for (a) to 0.46~dex.
 This is because, as shown in Table~\ref{tab:bl}, in the field range of the sample the enhancement by BL decreases with voltage more strongly at lower temperatures and differs between temperatures.
 \item \textbf{When $\mtun(E)$ is included, the presence or absence of BL cannot be distinguished.}
 For the nominal $N_\mathrm{D}$, (c) has the smaller RMSE and the smaller mean deviation for each temperature ($-0.06$ to $+0.05$~dex).
 The voltage dependence, on the other hand, is better reproduced by (d), whereas the residual change of (c) reaches $+0.6$~dex at low temperatures (Table~\ref{tab:perT}).
 Changing $N_\mathrm{D}$ by only $\pm10\%$ reverses their ranking.
 The effect of adding BL is effectively absorbed by an increase of $\phib$ by about 0.035~eV.
 \item \textbf{The obtained barrier heights are reasonable.}
 $\phib$ is 0.964~eV for (c) and 0.999~eV for (d), within the range between the value from the previous reverse analysis (0.96~eV) and
 the value from the forward characteristics (1.0~eV)\cite{Hatakeyama2003}.
 \item \textbf{The remaining discrepancy is at low temperatures.} At 30 and 40\degC, the RMSE of all models (a)--(d) exceeds 0.13~dex.
\end{enumerate}
Without punch-through in the interface field (second line of Eq.~(\ref{eq:Fs})), the RMSEs of (a), (b), and (d) increased by a factor of about 1.4--1.6,
whereas that of (c) decreased slightly from 0.109 to 0.093~dex (Table~\ref{tab:sens}).
Since punch-through necessarily occurs above about 450~V for this sample structure (Appendix~C),
this decrease results from the excessive voltage dependence of (c) (Table~\ref{tab:perT}) being compensated by an underestimate of the field at high voltage.

\subsection{Sensitivity to the sample parameters}
\label{sec:sensitivity}
Results for $N_\mathrm{D}$ changed by $\pm10\%$ were given in Table~\ref{tab:fit}.
The ranking of (c) and (d) depends on $N_\mathrm{D}$, and for $N_\mathrm{D}=5.5\times10^{15}\,\mathrm{cm^{-3}}$ the RMSE of (a) is also comparable to that of (d).
In the main text we mainly use the results for nominal values based on the design and Ref.~\citen{Hatakeyama2003}.
Table~\ref{tab:sens} shows the results for other conditions.
\begin{table}[htbp]
 \caption{\label{tab:sens}Fitting results under modified conditions: $\phib$ (eV) / RMSE (dex). Reference: $N_\mathrm{D}=5\times10^{15}\,\mathrm{cm^{-3}}$, $d=10\,\mu$m, with punch-through.}
 \footnotesize
 \begin{indented}
 \item[]\begin{tabular}{@{}lcccc}
 \br
 Condition & (a) & (b) & (c) & (d)\\
 \mr
 Reference & 0.949 / 0.166 & 0.987 / 0.241 & 0.964 / 0.109 & 0.999 / 0.120\\
 No punch-through & 0.943 / 0.263 & 0.982 / 0.331 & 0.957 / 0.093 & 0.993 / 0.177\\
 $d=9\,\mu$m & 0.954 / 0.113 & 0.993 / 0.183 & 0.971 / 0.203 & 1.006 / 0.161\\
 $d=11\,\mu$m & 0.945 / 0.213 & 0.984 / 0.286 & 0.960 / 0.072 & 0.996 / 0.137\\
 $N_\mathrm{D}=7\times10^{15}\,\mathrm{cm^{-3}}$ & 0.989 / 0.125 & 1.027 / 0.109 & 1.013 / 0.383 & 1.046 / 0.291\\
 \br
 \end{tabular}
 \end{indented}
\end{table}
Model (c) is the most sensitive to the conditions.
Under conditions that weaken the interface field at high voltage (no punch-through, $d=11\,\mu$m), the RMSE of (c) is 0.07--0.09~dex, the smallest of the four models,
whereas under conditions that strengthen the field ($d=9\,\mu$m, $N_\mathrm{D}=7\times10^{15}\,\mathrm{cm^{-3}}$) it is the largest, 0.20--0.38~dex.
The constant-mass models (a) and (b) show the opposite trend, and (d) is intermediate (because the image force partly cancels the increase in field dependence due to $\mtun(E)$).
$\mtun(E)$ decreases the mass for deeper tunneling and strengthens the field dependence of the current (Sect.~\ref{sec:consequence}).
It is therefore difficult to separate uniquely the error in the estimated interface field from the energy dependence of the mass using the present current characteristics alone.
The comparison with measurements above uses $N_\mathrm{D}=5\times10^{15}\,\mathrm{cm^{-3}}$ and $d=10\,\mu$m, based on the design and Ref.~\citen{Hatakeyama2003}, as the reference. Unless the sample parameters are evaluated independently, the relative merit of the mass models must be interpreted with this dependence on conditions in mind.

Note that the present model does not include the quantum-mechanical reflection at the metal--semiconductor interface (the interface-matching factor, i.e., the effective Richardson constant and its field dependence). Because this can change the field dependence of the calculated current, the relative merit of the mass models also depends on this treatment.

For reference, optimizing a constant mass together with the barrier height (triangular barrier without image force) gives $\mtun=0.265m_0$, $\phib=0.969\,\mathrm{eV}$, and an RMSE of 0.078~dex.
This fit is of a different kind from those of models (a)--(d). When the leakage current of an unknown SBD is to be predicted, the barrier height can be estimated to some extent in advance (e.g., from forward characteristics or internal photoemission), whereas the tunneling mass cannot in principle be set a priori.
A model in which the mass is also fitted therefore has no predictive basis, and its smaller RMSE has only reference value. The physical models in this paper are those in which only $\phib$ is fitted.
The fitted constant mass is a phenomenological parameter; its value lies within the range of the equivalent mass obtained from $\mtun(E)$ for the conditions of the sample ($0.25$--$0.32m_0$; Fig.~\ref{fig:meff}),
consistent with the interpretation that a constant mass is an equivalent mass determined by the depth of the current-carrying electrons (Sect.~\ref{sec:meff_interp}).

\section{Discussion}
\label{sec:discussion}

\subsection{Interpretation of effective masses obtained with a constant mass}
\label{sec:meff_interp}
\begin{figure}[htbp]
 \centering
 \includegraphics[width=\linewidth]{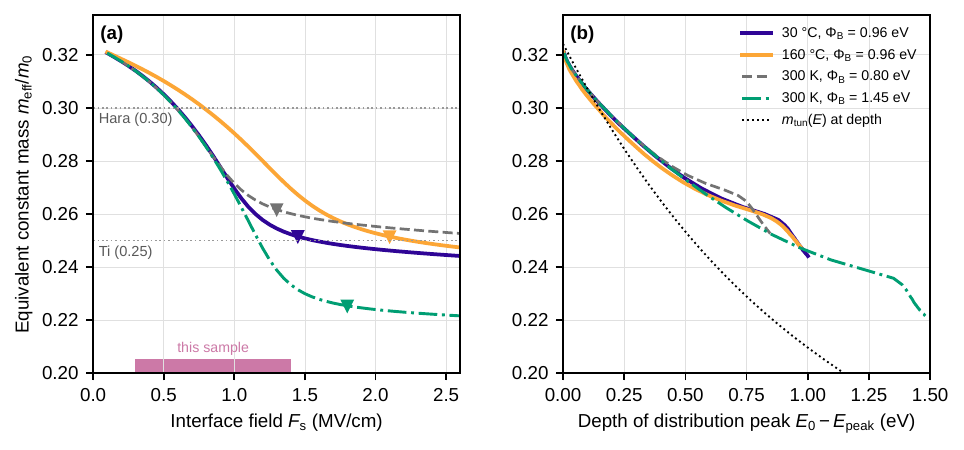}
 \caption{Equivalent constant mass $m_\mathrm{eff}$ that gives the same tunneling current through a triangular barrier as $\mtun(E)$ (Eq.~(\ref{eq:mE})).
 (a) Dependence on the interface field for 30 and 160\degC\ ($\phib=0.96\,\mathrm{eV}$; colors follow the temperature scale of Fig.~\ref{fig:fit})
 and for 300~K with $\phib=0.80\,\mathrm{eV}$ (gray dashed) and $1.45\,\mathrm{eV}$ (green dash-dotted; barrier height of Hara et al.\cite{Hara2022}).
 Triangles mark $\Fc$ for each condition ($\mtun=0.31m_0$), dotted lines show previously reported constant masses, and the band at the bottom shows the interface-field range of the sample. The equivalent masses under the conditions of Hara et al.\ are given in Table~\ref{tab:hara_meff}.
 (b) The same results plotted against the depth of the peak of the emission distribution, $E_0-E_\mathrm{peak}$. The black dotted line is $\mtun(E)$ at that depth.
 The TE component is excluded, and the image force is not included.}
 \label{fig:meff}
\end{figure}

We examine values obtained with constant-mass TFE formulas ($0.25m_0$ for Ti electrodes\cite{Hatakeyama2003,Okino2017} and $0.30m_0$ from fitting the forward and reverse characteristics of heavily doped samples with Ni electrodes\cite{Hara2022})
as equivalent masses of $\mtun(E)$.
Figure~\ref{fig:meff}(a) shows the constant mass $m_\mathrm{eff}$ that gives the same tunneling current through a triangular barrier as $\mtun(E)$ of Eq.~(\ref{eq:mE})
(the TE component, which does not depend on the mass, is excluded).
$m_\mathrm{eff}$ is close to the band-edge value $0.32m_0$ at low fields, decreases with increasing field,
and becomes nearly constant for $F_\mathrm{s}\gtrsim\Fc$, where the current-carrying electrons move to near the Fermi level.
For the conditions of the sample (0.3--1.4~MV/cm, 30--160\degC), $m_\mathrm{eff}=0.25$--$0.32m_0$,
and the previously reported value of $0.25m_0$ for Ti electrodes corresponds to the high-field, low-temperature end of this range.

Over the range of temperatures and barrier heights in Fig.~\ref{fig:meff}(b), $m_\mathrm{eff}$ follows a nearly common curve as a function of the depth of the peak of the emission distribution, $E_0-E_\mathrm{peak}$.
$m_\mathrm{eff}$ is larger than the local mass $\mtun(E)$ at that depth;
it is an effective quantity that includes the contributions of the tunneling path and the emission energy distribution, not a simple arithmetic mean of the local mass.
A constant mass can therefore be interpreted as an equivalent mass associated with the depth of the current-carrying electrons. The depth of the peak obtained numerically here must, however, be distinguished from $\Delta$ given by the constant-mass saddle-point formula (Sect.~\ref{sec:index}).
Part of the variation in reported effective masses may reflect differences in the analyzed field range and temperature (i.e., in $F_\mathrm{s}/\Fc$),
and the difference between electrode metals\cite{Okino2017} may also include a difference in $\Fc$ through the barrier height. Treu et al.\cite{Treu2001} likewise obtained $0.19$--$0.30m_0$ from the reverse characteristics of W, Ti, and Ta electrodes ($\phib=0.95$--$1.27\,\mathrm{eV}$).

The value $0.30m_0$ of Hara et al.\cite{Hara2022} was obtained by fitting the forward and reverse current--voltage characteristics of Ni/4H-SiC ($N_\mathrm{D}=4.2\times10^{15}$--$1.2\times10^{19}\,\mathrm{cm^{-3}}$)
with the barrier height and a constant mass as parameters.
Table~\ref{tab:hara_meff} shows the constant mass that gives the same current as $\mtun(E)$ of Eq.~(\ref{eq:mE}) in their barrier model (quadratic barrier with image force, same conditions as Table~\ref{tab:hara}).
At the reverse TFE/FE transition points, $m_\mathrm{eff}=0.23m_0$, comparable to the value for $F_\mathrm{s}\gtrsim\Fc$ in Fig.~\ref{fig:meff}(a).
In forward bias, by contrast, the tunneling depth varies with $N_\mathrm{D}$ and forward voltage, and $m_\mathrm{eff}=0.25$--$0.32m_0$.
Because the forward data cover all samples, a single constant mass fitted simultaneously to forward and reverse characteristics is likely to fall in this range, and the reported $0.30m_0$ does.
The reported value is therefore consistent with $\mtun(E)$. Conversely, $\mtun(E)$ predicts that a constant-mass fit restricted to the vicinity of the reverse transition points would yield about $0.23m_0$.
Since the weights of the forward and reverse data in their fit cannot be determined quantitatively from the original paper, this comparison is kept qualitative.

\begin{table}[htbp]
 \caption{\label{tab:hara_meff}Equivalent constant mass $m_\mathrm{eff}$ (in units of $m_0$) that gives the same current as $\mtun(E)$ (Eq.~(\ref{eq:mE})) under the conditions of Hara et al.\cite{Hara2022}.
 Quadratic barrier with image force, WKB, Fermi--Dirac supply, $T=300\,\mathrm K$, $\epss=10.32\varepsilon_0$.
 Reverse bias: transition points of Table~\ref{tab:hara} ($V_\mathrm{R}$ determined with $\phib=1.43$--$1.48\,\mathrm{eV}$ and $\mtun=0.30m_0$); forward bias: $\phib=1.45\,\mathrm{eV}$.}
 \begin{indented}
 \item[]\begin{tabular}{@{}lccc}
 \br
 & \multicolumn{3}{c}{$N_\mathrm{D}$ (cm$^{-3}$)}\\
 Condition & $2.0\times10^{17}$ & $2.6\times10^{18}$ & $1.2\times10^{19}$\\
 \mr
 Reverse, TFE/FE transition & 0.225--0.227 & 0.227--0.229 & 0.231--0.233\\
 Forward, $V_\mathrm{F}=0.2\,\mathrm V$ & 0.318 & 0.287 & 0.254\\
 Forward, $V_\mathrm{F}=0.5\,\mathrm V$ & 0.319 & 0.296 & 0.270\\
 Forward, $V_\mathrm{F}=0.8\,\mathrm V$ & 0.320 & 0.305 & 0.288\\
 \br
 \end{tabular}
 \end{indented}
\end{table}

On the other hand, $0.15m_0$ for Ni electrodes and $0.20m_0$ for Mo electrodes\cite{Okino2017} are below the lower limit of $m_\mathrm{eff}$ in the present model ($0.22m_0$ within the range of the figure).
These differences likely include factors other than the mass, such as differences in analysis method (determination of the barrier height and of the TFE--FE transition), interface states, and barrier-height inhomogeneity.

\subsection{Distinguishability of the image force}
Among the four models compared, model (d) includes both the image force and the energy-dependent mass.
In the present data, however, the effect of BL is nearly indistinguishable from a change of about 0.035~eV in $\phib$.
For the constant-mass model in Table~\ref{tab:bl}, the enhancement of the total current over 0.3--1.4~MV/cm is about 2.7--4.6 at 300~K and about 3.0--3.8 at 433~K. This result is a guide for understanding the action of the image force and does not directly represent the difference between the fits of models (c) and (d).
In the fits, most of the current increase due to the image force is compensated by an increase of $\phib$, so its contribution cannot be separated from the reproducibility of the present data alone.
Identifying BL requires high-precision measurements on the low-field side, where the enhancement changes strongly, or a combination with measurements that determine $\phib$ independently (e.g., internal photoemission).
That lowering only the TE barrier height with a constant mass is insufficient for evaluating the tunneling current agrees with the observation of Li et al.\cite{Li2022}.

\subsection{Discrepancy at low temperatures}
At 30 and 40\degC, the discrepancy is large for all models (a)--(d) compared in Sect.~\ref{sec:exp}. At 30\degC, $F_\mathrm{s}/\Fc=0.96$ at 950~V, close to the TFE/FE boundary where the saddle point of the reference model approaches the Fermi level (Fig.~\ref{fig:regime}).
In this region, the use of $\mtun(E)$ of Eq.~(\ref{eq:mE}) deep in the gap,
the treatment of the WKB approximation for a position-dependent mass, and the accuracy of the electron supply model all become critical.
On the low-voltage side, the current is about $10^{-8}\,\mathrm{A/cm^2}$, close to the measurement floor, and noise also contributes.
Lateral inhomogeneity of the barrier height\cite{WernerGuttler1991,Tung1992} can also affect the current more at lower temperatures.
In addition, current components other than the current through the Schottky barrier are possible: generation--recombination current in the depletion region (bulk), surface generation--recombination current at the oxide interface of the termination, and electron emission from deep levels in the bulk.
Their temperature and field dependences differ from those of the current through the Schottky barrier, and they may not be negligible at low temperatures, where the current is small.
Separating these factors remains for future work.

\subsection{Excess current at high voltages}
Above 950~V, the measured current exceeds the calculation, increases sharply from about 1050~V, and reaches the current limit above 1100~V (Fig.~\ref{fig:fit}).
Estimating the one-dimensional multiplication in the active region with the impact ionization coefficients of 4H-SiC\cite{Hatakeyama2004} gives
$M_\mathrm{n}-1\approx6\times10^{-7}$ (950~V), $8\times10^{-6}$ (1100~V), and $1\times10^{-4}$ (1300~V),
so avalanche in the active region cannot explain this excess current.
Breakdown at the edge termination (JTE), where the field is concentrated, is a candidate, but its contribution cannot be established from the present data alone.
In addition, the current density at 950~V and 160\degC\ is about 0.17~A/cm$^2$, giving a power density above 100~W/cm$^2$;
in DC measurements, self-heating may raise the junction temperature above the set temperature and increase the leakage current.
The contribution of the termination may be predicted by TCAD simulations including the JTE and impact ionization.
However, simulating the whole device is not feasible in terms of computational cost; only part of the termination can be simulated and the result scaled by the perimeter, so the excess current cannot necessarily be evaluated quantitatively.
The present model, including the energy-dependent mass, can be incorporated into TCAD (Sect.~\ref{sec:tcad}).
In new measurements, self-heating should be suppressed by pulsed measurements.
In addition, the contribution of the termination should be evaluated from the differences in leakage current among devices of a test element group (TEG) that contains devices with different termination structures.
Including devices with different electrode diameters further allows the leakage current to be divided into components proportional to the area and to the perimeter, which separates the contribution of the periphery.

\subsection{Implementation in device simulation}
\label{sec:tcad}
\begin{figure}[htbp]
 \centering
 \includegraphics[width=\linewidth]{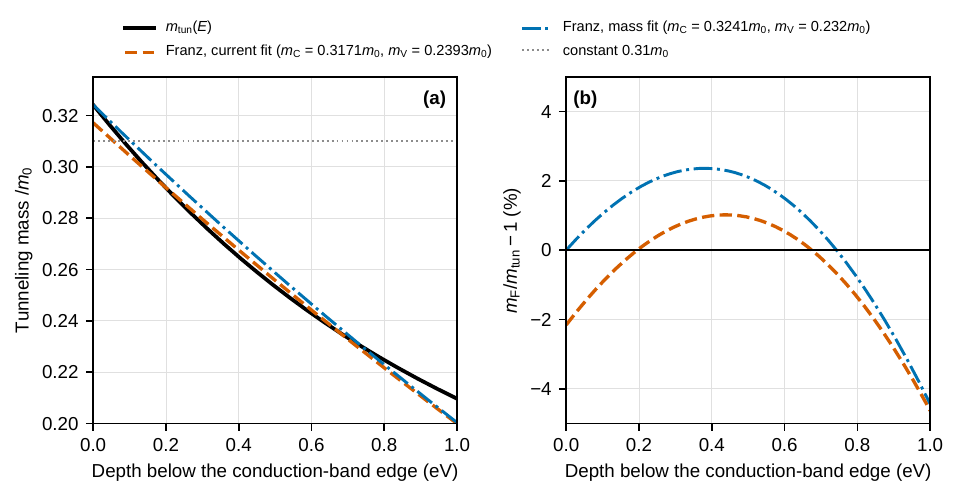}
 \caption{(a) Energy-dependent tunneling mass $\mtun(E)$ of Eq.~(\ref{eq:mE}) (solid) and the Franz two-band mass $m_\mathrm{F}$ of Eq.~(\ref{eq:franz}) as functions of the depth below the conduction-band edge, $-\mathcal{E}$.
 Dashed: $m_\mathrm{C}=0.3171m_0$, $m_\mathrm{V}=0.2393m_0$ fitted to the WKB current (used in the simulations); dash-dotted: $m_\mathrm{C}=0.3241m_0$, $m_\mathrm{V}=0.232m_0$ fitted to $\mtun(E)$ itself over 0--1~eV. The dotted line is the constant mass $0.31m_0$. $E_\mathrm{g}=3.26\,\mathrm{eV}$.
 (b) Relative difference $m_\mathrm{F}/\mtun-1$.}
 \label{fig:franz}
\end{figure}

The present model can be used in a commercial device simulator.
In Sentaurus Device\cite{SentaurusUG}, the nonlocal barrier-tunneling model computes the WKB transmission from a Schottky contact with the same current integral as Eq.~(\ref{eq:J}),
but the tunneling mass is a constant and cannot be replaced through the physical model interface.
With the TwoBand option, however, the imaginary wave number in the gap follows the Franz two-band dispersion\cite{Franz1956}, $\kappa^{-2}=\kappa_\mathrm{C}^{-2}+\kappa_\mathrm{V}^{-2}$,
which is equivalent to the energy-dependent mass
\begin{equation}
 m_\mathrm{F}(\mathcal{E})=\frac{m_\mathrm{C}m_\mathrm{V}(E_\mathrm{g}+\mathcal{E})}{m_\mathrm{V}(E_\mathrm{g}+\mathcal{E})-m_\mathrm{C}\mathcal{E}},
 \label{eq:franz}
\end{equation}
with $\mathcal{E}$ of Eq.~(\ref{eq:mE}) ($-\mathcal{E}$ is the depth below the conduction-band edge). Sentaurus Device also provides the Kane two-band dispersion\cite{Kane1961} for band-to-band tunneling, in which $m_\mathrm{C}$ and $m_\mathrm{V}$ are tied to a reduced mass; in the Franz form used here they are independent.
Fitting the WKB currents with $\mtun(E)$ of Eq.~(\ref{eq:mE}) for 0.3--1.4~MV/cm and 30--160\degC\ gives $m_\mathrm{C}=0.3171m_0$ and $m_\mathrm{V}=0.2393m_0$
($m_\mathrm{V}$ is a fitting parameter, not the hole mass), with an rms difference of 0.005~dex.
Figure~\ref{fig:franz} compares the resulting $m_\mathrm{F}$ with $\mtun(E)$. The difference is within 2.2\% down to 0.8~eV below the conduction-band edge and 4.6\% at 1~eV.
A direct fit of $m_\mathrm{F}$ to $\mtun(E)$ over 0--1~eV gives $m_\mathrm{C}=0.3241m_0$ and $m_\mathrm{V}=0.232m_0$ with a maximum mass difference of 4.4\%, but the current difference then reaches 0.057~dex; the values fitted to the current are therefore used.

One-dimensional simulations of the sample of Sect.~\ref{sec:exp} with these masses, the barrier height of model (c), and the same $\Ar$, $N_\mathrm{C}$, and $\epss$ (supplementary data, Sect.~S7)
reproduce the WKB currents within 0.013~dex at 950~V at all temperatures.
With the image force (\texttt{BarrierLowering}, model (d)), the agreement at 950~V is within 0.035~dex, provided that the upper limit of the image-force lowering, which is 0.3~eV by default, is raised.
At low voltages the simulated currents are smaller, by up to 0.10~dex (0.13~dex with the image force) at 50~V and 160\degC.
This difference arises from the Schottky boundary condition of the drift-diffusion model: without tunneling, the simulated TE current is 0.77 times $\Ar T^2e^{-\phib/\kb T}$, independent of voltage,
which corresponds to the thermionic-diffusion factor $1/(1+v_\mathrm{R}/v_\mathrm{D})$\cite{CrowellSze1966} with $v_\mathrm{R}/v_\mathrm{D}\approx0.3$, i.e., less than 10~meV in barrier height.
With the Franz two-band dispersion, the energy-dependent tunneling mass can thus be incorporated directly into a commercial device simulator.
In JBS structures, the field at the Schottky interface between the p$^+$ regions varies two-dimensionally with the cell geometry and the reverse voltage, and the leakage current depends strongly on this interface field.
Through TCAD, the emission regimes and energy scales summarized in this paper, the tunneling mass based on the complex band structure, and the validation against measured data can be used to predict the leakage current of such structures and thus applied to the design of JBS structures.

\section{Conclusions}
\label{sec:conclusion}
We treated the reverse leakage current of 4H-SiC SBDs as an energy integral of Fermi--Dirac supply and WKB transmission and organized the emission mechanisms and tunneling transport theoretically.
For a triangular barrier with constant mass and without image force, $F_\mathrm{s}/\Fc$ gives the saddle depth $\Delta=\phib(F_\mathrm{s}/\Fc)^2$, and the saddle point reaches the Fermi level at $F_\mathrm{s}=\Fc$.
In the non-punch-through depletion approximation, $F_\mathrm{s}/\Fc=(E_{00}/\kb T)\sqrt{B/\phib}$, which explains why tunneling becomes important at high reverse bias even in low-doped layers with small $E_{00}/\kb T$.
The Airy energy $E_\mathrm{A}$ and Airy length $l_\mathrm{A}$ are the scales that govern the decay of the electron wave function in the barrier and follow from the same Airy equation as the two-dimensional electrons in a triangular well.
Using these scales, we distinguished the crossover between the TE and tunneling components ($E_\mathrm{A}\approx0.80\kb T$),
the guide for the TFE saddle-point approximation ($\Delta=3\kb T$), and the formal TFE/FE boundary ($\Delta=\phib$).
The TFE/FE boundary generalizes to $c_1\kb T\approx1$ with the slope of the exponent at the Fermi level, and at the transition points of heavily doped layers $c_1\kb T\approx0.87$ independently of $N_\mathrm{D}$ up to $1.2\times10^{19}\,\mathrm{cm^{-3}}$.

We summarized closed-form expressions for the reverse FE and TFE currents of triangular (FN, FE, and TFE) and quadratic (FE and TFE) barriers and re-examined the PS formulas.
The PS reverse FE formulas, Eqs.~(37)--(39) of PS, are the leading terms in the limit of a reverse voltage much larger than the barrier height and are effectively the triangular-barrier formula. In heavily doped layers, where the curvature of the barrier is not negligible ($N_\mathrm{D}=10^{19}\,\mathrm{cm^{-3}}$, 1--10~V), they overestimate the WKB current of the quadratic barrier 1.4--7.9-fold, whereas the FE expression with the exact $b_1$ and $c_1$ reproduces it within 15\%.
For low-doped drift layers at 300--433~K (the conditions of Table~\ref{tab:app_ps}), the PS reverse TFE formula, Eq.~(50) of PS, gives 2.5--3.6 times the total WKB current because of the approximated curvature and the one-sided Gaussian integral in its prefactor; under the same conditions, the expression with the exact curvature at the saddle point agrees with the total WKB current within about 2\%. Under the conditions of low-doped layers, the saddle-point approximation for the quadratic barrier reduces to the triangular-barrier TFE formula.
The image force changes the action integral in addition to lowering the barrier top; for a constant mass at 300~K, the enhancement of the total current is largest near 0.5~MV/cm and decreases at higher fields, and the position of this maximum also depends on temperature.
The energy-dependent tunneling effective mass, on the other hand, captures the effect of deep energies involved at high fields and low temperatures.
This view provides a basis for interpreting conventional constant masses as equivalent masses that depend on the analyzed range. For example, $0.30m_0$ of Hara et al.\ is consistent with the equivalent masses of $\mtun(E)$ under the same conditions ($0.25$--$0.32m_0$ in forward bias and $0.23m_0$ at the reverse transition points).

As an application, we reanalyzed the reverse characteristics of a Ti/4H-SiC SBD at 30--160\degC\ and 50--950~V.
In a comparison of four models fitted with a single barrier height common to all temperatures, introducing $\mtun(E)$ under the nominal sample parameters gave an RMSE of 0.11--0.12~dex without a mass-adjustment parameter and improved the reproduction of the voltage dependence.
The donor concentration and layer thickness are correlated with the mass model, and the image-force contribution cannot be separated from an adjustment of the barrier height by about 0.035~eV; in addition, the model does not include the quantum-mechanical reflection at the interface. Independent sample characterization is therefore needed to identify each effect.
The characteristics at low temperatures and high voltages will be examined further by measurements that separate current components other than the current through the Schottky barrier (deep levels, surface, and termination) and self-heating.

The predictive physical model is the one in which only $\phib$ is fitted and the energy dependence of the mass is taken from the band structure, because the tunneling mass cannot be set in advance for an unknown device; a constant mass fitted together with the barrier height (RMSE 0.078~dex) is given for reference.
This model can be implemented in Sentaurus Device with the TwoBand option, which uses the Franz two-band dispersion: two masses fitted to the current reproduce the current with $\mtun(E)$ within 0.005~dex (rms) and approximate the mass within 2.2\% down to 0.8~eV below the conduction-band edge. The model can therefore be used to predict the leakage current in the design of JBS structures.

\section*{Data availability statement}
The code that reproduces the figures and tables of this paper and the supplementary data, the measured reverse current--voltage data, and the Sentaurus Device input files are openly available at \url{https://doi.org/10.5281/zenodo.24337950}.

\ack
This work was supported by JSPS KAKENHI Grant Numbers JP23K03928 and JP26K07588, by Synopsys Japan G.K. through the activities of VDEC, d.lab, The University of Tokyo, and by an academic research grant from Toshiba Electronic Devices \& Storage Corporation.
The author is deeply grateful to Prof.~Shin-Ichiro Kuroki, Prof.~Manabu Arai, and Prof.~Akira Toriumi for providing the opportunity for this research.

\appendix
\section{TFE formula for the triangular barrier (Hatakeyama formula) and the critical field}
\label{app:tri}
This appendix derives the triangular-barrier TFE formula~(\ref{eq:tfe_tri_field_equiv}) of Sect.~\ref{sec:closed} (Eq.~(3) of Hatakeyama et al.\cite{HatakeyamaShinohe2002}) and the critical field $\Fc$.
When the supply function is approximated by the Boltzmann form well above the Fermi level, the integrand of Eqs.~(\ref{eq:J}) and (\ref{eq:Gtri}) becomes $\exp[-g(E)]$ with
\begin{equation}
 g(E)=\frac{4\sqrt{2\mtun}}{3e\hbar F_\mathrm{s}}(E_0-E)^{3/2}+\frac{E-\Ef}{\kb T}.
\end{equation}
From $g'(E_\mathrm{p})=0$, the saddle point is
\begin{equation}
 E_0-E_\mathrm{p}=\left(\frac{e\hbar F_\mathrm{s}}{2\sqrt{2\mtun}\kb T}\right)^2=\phib\left(\frac{F_\mathrm{s}}{\Fc}\right)^2,\qquad
 \Fc=\frac{2\kb T\sqrt{2\mtun\phib}}{e\hbar},
 \label{eq:tfe_fc}
\end{equation}
and the value and curvature at the saddle point are
\begin{equation}
 g(E_\mathrm{p})=\frac{\phib}{\kb T}-\frac{(e\hbar F_\mathrm{s})^2}{24\mtun(\kb T)^3},\qquad
 g''(E_\mathrm{p})=\frac{4\mtun\kb T}{(e\hbar F_\mathrm{s})^2}.
\end{equation}
Performing the Gaussian integral $\int e^{-g''(E-E_\mathrm{p})^2/2}dE=\sqrt{2\pi/g''}$ gives Eq.~(\ref{eq:tfe_tri_field_equiv}).
The second term in the exponent is the increment due to tunneling and equals $\phib(F_\mathrm{s}/\Fc)^2/(3\kb T)$.
With the Airy energy $E_\mathrm{A}=[(e\hbar F_\mathrm{s})^2/(2\mtun)]^{1/3}$ and $e\hbar F_\mathrm{s}/\sqrt{2\mtun}=E_\mathrm{A}^{3/2}$,
\begin{equation}
 \Jtfe=\Ar T^2\exp\left(-\frac{\phib}{\kb T}\right)\cdot\sqrt\pi\left(\frac{E_\mathrm{A}}{\kb T}\right)^{3/2}\exp\left[\frac{1}{12}\left(\frac{E_\mathrm{A}}{\kb T}\right)^3\right].
\end{equation}
The first factor is the TE current, so the TFE current is the TE current multiplied by a function of $E_\mathrm{A}/\kb T$ alone.

The formula requires the saddle point to lie above the Fermi level ($F_\mathrm{s}<\Fc$) with $E_\mathrm{p}-\Ef\gg\kb T$,
and to lie more than about the standard deviation of the Gaussian, $\sigma_E=1/\sqrt{g''}=\sqrt{2\kb T\Delta}$, below the barrier top, i.e., $\Delta\gtrsim2\kb T$ (the limit on the low-field side, where TFE approaches TE).
For $F_\mathrm{s}>\Fc$, the FE formula obtained by linearizing the exponent at the Fermi level is used; its convergence condition is exactly $F_\mathrm{s}>\Fc$
(Eq.~(\ref{eq:fe_tri}); the FE and FN formulas are derived in Sect.~S3 of the supplementary data).
The image force is not included in this formula. Inserting a barrier height lowered by $\Delta U$ into it is not used,
because it does not include the change of barrier shape (Sect.~\ref{sec:bl}).

\section{Re-evaluation of the Padovani--Stratton reverse FE and TFE formulas}
\label{app:ps}
We evaluate the FE and TFE currents for the quadratic depletion barrier and compare them with the reverse formulas of PS\cite{PadovaniStratton1966}.
In this appendix $\Ef=0$. Energies in the current formulas are in SI units (J), and the numerical table is given in eV; when energy ratios are evaluated in eV, all energies including $\kb T$ must be in the same unit.

\subsection{FE formula}
Expanding the exponent of Eq.~(\ref{eq:Gquad}) to first order about the Fermi level ($E=0$), $G(E)\simeq b_1-c_1E$, gives $b_1$ and $c_1$ of Eq.~(\ref{eq:B_b1c1}) and the FE current~(\ref{eq:B_fe}) (supplementary data, Sect.~S4.4).
Equations~(34) and (35) of PS agree with Eq.~(\ref{eq:B_b1c1}) for $eV_\mathrm{n}=0$.
Equations~(37) and (38) of PS are the leading terms for $\phib/X\ll1$, $b=2\phib^{3/2}/(3E_{00}\sqrt B)$ and $c=\sqrt{\phib/B}/E_{00}$, with $eV_\mathrm{n}=0$; these are $b$ and $c$ of the triangular barrier.
The ratios are $b_1/b=1+\phib/(5X)+\cdots$ and $c_1/c=1+\phib/(3X)+\cdots$.

\subsection{Saddle point and exponent}
With $X=eV_\mathrm{R}-eV_\mathrm{n}$, the bulk conduction-band edge is $E_{\mathrm{C,b}}=-X$ and the band bending is $B=\phib+X$.
The derivatives of the exponent of Eq.~(\ref{eq:Gquad}) are
\begin{equation}
 G'(E)=-\frac{1}{E_{00}}\mathrm{arcosh}\frac{1}{\sqrt u},\qquad
 G''(E)=\frac{1}{2E_{00}B\,u\sqrt{1-u}},\qquad u=\frac{E-E_{\mathrm{C,b}}}{B}.
 \label{eq:app_ps_dG}
\end{equation}
With Boltzmann supply, the saddle point where $G'(E_\mathrm{m})=-1/\kb T$ is, with $t=E_{00}/\kb T$,
\begin{equation}
 E_\mathrm{m}=-X+B\,\mathrm{sech}^2t,
\end{equation}
which lies $B\tanh^2t$ below the barrier top. The exponent at the saddle point is
\begin{equation}
 G(E_\mathrm{m})+\frac{E_\mathrm{m}}{\kb T}=\frac{B}{E_{00}}\tanh t-\frac{X}{\kb T}
 =\frac{\phib}{E_{0,\mathrm{PS}}}-\frac{X}{E'}
\end{equation}
with $E_{0,\mathrm{PS}}=E_{00}\coth t$ and $E'=E_{00}/(t-\tanh t)$.
This is the same as Eqs.~(45)--(49) of PS; the voltage slope $1/E'$ and the temperature dependence agree with their result.

\subsection{Prefactor}
The curvature at the saddle point is $G''(E_\mathrm{m})=\cosh^2t/(2E_{00}B\tanh t)$. Performing the Gaussian integral $\int e^{-G''(E-E_\mathrm{m})^2/2}dE=\sqrt{2\pi/G''}$ gives Eq.~(\ref{eq:app_ps_tfe}).
The prefactor of Eq.~(50) of PS is $(\Ar T/\kb)\sqrt{\pi E_{00}(eV_\mathrm{R}+\phib\,\mathrm{sech}^2t)}$, and the ratio of the two is
\begin{equation}
 \frac{J_\mathrm{PS(50)}}{\Jtfe^\mathrm{quad}}
 =\frac12\cdot\frac{1}{\sqrt{\tanh t}}\cdot
 \sqrt{\frac{eV_\mathrm{R}+\phib\,\mathrm{sech}^2t}{B\,\mathrm{sech}^2t}}.
 \label{eq:app_ps_ratio}
\end{equation}
The second factor is the main difference. The curvature coefficient of PS (their Eqs.~(24) and (44)) corresponds to setting $\sqrt{1-u}$ in Eq.~(\ref{eq:app_ps_dG}) to unity,
which is an approximation valid when the saddle point lies well below the barrier top ($u\ll1$).
In the TFE regime, particularly for $E_{00}\ll\kb T$, $u=\mathrm{sech}^2t\simeq1$, so the curvature is underestimated by a factor of $\tanh t$ and the current is overestimated.
The first factor arises because Eq.~(50) of PS corresponds to one side of the Gaussian integral (presumably corresponding to the omission of the error-function term in their Eq.~(17)).
The third factor reflects the difference in the bracket and is nearly unity for $t\ll1$.
Since the exponents agree, the effect on the slopes and barrier heights used by PS in their analysis of experiments is small.

\subsection{Numerical check and relation to the triangular barrier}
Table~\ref{tab:app_ps} shows the errors of Eq.~(\ref{eq:app_ps_tfe}) and of Eq.~(50) of PS relative to the WKB numerical integral~(\ref{eq:J}) for the quadratic barrier.
For these conditions, Eq.~(\ref{eq:app_ps_tfe}) agrees with the total WKB current, including TE, within about 2\% (within 8\% for the tunneling component alone),
whereas Eq.~(50) of PS gives currents 2.5--3.6 times larger for the SiC drift layers at 300--433~K, 1.45 times larger at 100~K, and 14\% smaller for $10^{18}\,\mathrm{cm^{-3}}$ (1~V, outside the range of validity of PS Eq.~(51)).
The error of Eq.~(\ref{eq:app_ps_tfe}) increases when the saddle point approaches the barrier top within about the standard deviation of the Gaussian, $\sigma_E=1/\sqrt{G''(E_\mathrm{m})}$; for example, it is $-6.6\%$ for $10^{16}\,\mathrm{cm^{-3}}$, 300~K, and 10~V, and reaches about 15\% at the edge of the range of validity (Fig.~S2).
\begin{table}[htbp]
 \caption{\label{tab:app_ps}Errors of the TFE formulas relative to the total WKB current, Eq.~(\ref{eq:J}), including the TE component above the barrier top. $\phib=1.0\,\mathrm{eV}$, $\mtun=0.31m_0$, $\varepsilon_\mathrm{r}=10.03$. $E_\mathrm{m}$ is the saddle point measured from the Fermi level. $^\dagger$Below the minimum bias for TFE given by PS Eq.~(51) (about 1.25~V for this condition), i.e., outside the range of validity of the PS formula.}
 \begin{indented}
 \item[]\begin{tabular}{@{}rrrrrrr}
 \br
 $N_\mathrm{D}$ ($\mathrm{cm^{-3}}$) & $T$ (K) & $V_\mathrm{R}$ (V) & $t$ & $E_\mathrm{m}$ (eV) & Eq.~(\ref{eq:app_ps_tfe}) & PS Eq.~(50)\\
 \mr
 $5\times10^{15}$ & 300 & 300 & 0.029 & 0.751 & $+1.1\%$ & $+198\%$\\
 $5\times10^{15}$ & 433 & 300 & 0.020 & 0.880 & $+2.0\%$ & $+261\%$\\
 $10^{16}$ & 100 & 10 & 0.122 & 0.838 & $+0.5\%$ & $+45\%$\\
 $10^{16}$ & 300 & 100 & 0.041 & 0.833 & $+1.5\%$ & $+152\%$\\
 $10^{18}$ & 300 & 1 & 0.407 & 0.713 & $+1.1\%$ & $-14\%^\dagger$\\
 \br
 \end{tabular}
 \end{indented}
\end{table}

Expanding $\tanh t\simeq t-t^3/3$ for $t\ll1$ and using $F_\mathrm{s}^2=2N_\mathrm{D}B/\epss$ ($B$ in J), Eq.~(\ref{eq:app_ps_tfe}) agrees with
the triangular-barrier TFE formula~(\ref{eq:tfe_tri_field_equiv}) in both exponent and prefactor.
The next term in the exponent is $-(2/15)(B/\kb T)t^4$; for $N_\mathrm{D}=5\times10^{15}\,\mathrm{cm^{-3}}$, $V_\mathrm{R}\leq300\,\mathrm V$, and 30--160\degC,
the two formulas differ by less than 0.3\%. Under these conditions, therefore, the triangular-barrier TFE formula is practically equivalent to the quadratic-barrier TFE formula
and more accurate than Eq.~(50) of PS.

\section{Numerical methods}
\label{app:num}
\subsection{Interface field and built-in potential}
\label{app:field}
The Fermi-level position in the neutral region was $eV_\mathrm{n}=\kb T\ln[\Nc(T)/N_\mathrm{D}]$ with the effective density of states
$\Nc(T)=1.7\times10^{19}(T/300)^{3/2}\,\mathrm{cm^{-3}}$, and $eV_\mathrm{bi}=\phib-eV_\mathrm{n}$.
When the depletion width $W_\mathrm{d}=\sqrt{2\epss(V_\mathrm{R}+V_\mathrm{bi})/(eN_\mathrm{D})}$ does not exceed the drift-layer thickness $d$, the first line of Eq.~(\ref{eq:Fs}) was used;
for $W_\mathrm{d}>d$ (above about 450~V for the sample), the second line was used for the punch-through state in which the whole drift layer is depleted.
Depletion on the substrate side and the image-force correction to the interface field were neglected. We used the relative permittivity along the c-axis, $\varepsilon_\mathrm{r}=10.03$\cite{PatrickChoyke1970} (a value measured for 6H-SiC, which is commonly used for 4H-SiC as well).

\subsection{Fitting procedure}
\label{app:fit}
For each temperature, 37 evaluation voltages were set at 25~V intervals over 50--950~V, and the median of $\log_{10}J$ of the measured points within $\pm12.5\,\mathrm V$ of each evaluation voltage was used as the measured value.
The figure of merit was the RMSE of $\log_{10}(J_\mathrm{calc}/J_\mathrm{meas})$ over 296 points at eight temperatures,
and the RMSE was minimized with respect to $\phib$ over the range 0.55--1.15~eV using Brent's method.

The sensitivity analysis is given in Sect.~\ref{sec:sensitivity} and Table~\ref{tab:sens}.


\section*{References}
\begin{thebibliography}{99}
\bibitem{Kimoto2015} T. Kimoto, Jpn. J. Appl. Phys. \textbf{54}, 040103 (2015).
\bibitem{Bhatnagar1992} M. Bhatnagar, P. K. McLarty, and B. J. Baliga, IEEE Electron Device Lett. \textbf{13}, 501 (1992).
\bibitem{Hatakeyama2003} T. Hatakeyama, M. Kushibe, T. Watanabe, S. Imai, and T. Shinohe, Mater. Sci. Forum \textbf{433--436}, 831 (2003).
\bibitem{CroftonSriram1996} J. Crofton and S. Sriram, IEEE Trans. Electron Devices \textbf{43}, 2305 (1996).
\bibitem{Treu2001} M. Treu, R. Rupp, H. Kapels, and W. Bartsch, Mater. Sci. Forum \textbf{353--356}, 679 (2001).
\bibitem{Baliga2008} B. J. Baliga, \textit{Fundamentals of Power Semiconductor Devices} (Springer, New York, 2008).
\bibitem{Baliga1984} B. J. Baliga, IEEE Electron Device Lett. \textbf{5}, 194 (1984).
\bibitem{SzeNg2007} S. M. Sze and K. K. Ng, \textit{Physics of Semiconductor Devices} (Wiley, Hoboken, NJ, 2007) 3rd ed.
\bibitem{Rhoderick1988} E. H. Rhoderick and R. H. Williams, \textit{Metal--Semiconductor Contacts} (Clarendon Press, Oxford, 1988) 2nd ed.
\bibitem{HatakeyamaShinohe2002} T. Hatakeyama and T. Shinohe, Mater. Sci. Forum \textbf{389--393}, 1169 (2002).
\bibitem{FowlerNordheim1928} R. H. Fowler and L. Nordheim, Proc. R. Soc. Lond. A \textbf{119}, 173 (1928).
\bibitem{MurphyGood1956} E. L. Murphy and R. H. Good, Jr., Phys. Rev. \textbf{102}, 1464 (1956).
\bibitem{PadovaniStratton1966} F. A. Padovani and R. Stratton, Solid-State Electron. \textbf{9}, 695 (1966).
\bibitem{Stratton1962} R. Stratton, Phys. Rev. \textbf{125}, 67 (1962).
\bibitem{CrowellRideout1969} C. R. Crowell and V. L. Rideout, Solid-State Electron. \textbf{12}, 89 (1969).
\bibitem{Li2022} W. Li, D. Jena, and H. G. Xing, J. Appl. Phys. \textbf{131}, 015702 (2022).
\bibitem{Hara2022} M. Hara, H. Tanaka, M. Kaneko, and T. Kimoto, Appl. Phys. Lett. \textbf{120}, 172103 (2022).
\bibitem{Murakami2023} Y. Murakami, S. Nagamizo, H. Tanaka, and N. Mori, Jpn. J. Appl. Phys. \textbf{62}, SC1042 (2023).
\bibitem{TsuEsaki1973} R. Tsu and L. Esaki, Appl. Phys. Lett. \textbf{22}, 562 (1973).
\bibitem{Wellenhofer1997} G. Wellenhofer and U. R\"ossler, Phys. Status Solidi B \textbf{202}, 107 (1997).
\bibitem{Crowell1965} C. R. Crowell, Solid-State Electron. \textbf{8}, 395 (1965).
\bibitem{Crowell1969} C. R. Crowell, Solid-State Electron. \textbf{12}, 55 (1969).
\bibitem{CrowellSze1966} C. R. Crowell and S. M. Sze, Solid-State Electron. \textbf{9}, 1035 (1966).
\bibitem{Nicholls2019} J. Nicholls, S. Dimitrijev, P. Tanner, and J. Han, Sci. Rep. \textbf{9}, 3754 (2019).
\bibitem{PatrickChoyke1970} L. Patrick and W. J. Choyke, Phys. Rev. B \textbf{2}, 2255 (1970).
\bibitem{DLMF} F. W. J. Olver, A. B. Olde Daalhuis, D. W. Lozier, B. I. Schneider, R. F. Boisvert, C. W. Clark, B. R. Miller, B. V. Saunders, H. S. Cohl, and M. A. McClain (eds.), \textit{NIST Digital Library of Mathematical Functions, Chap. 9: Airy and Related Functions} [\url{https://dlmf.nist.gov/9}].
\bibitem{Stern1972} F. Stern, Phys. Rev. B \textbf{5}, 4891 (1972).
\bibitem{Ando1982} T. Ando, A. B. Fowler, and F. Stern, Rev. Mod. Phys. \textbf{54}, 437 (1982).
\bibitem{DLMF12} F. W. J. Olver, A. B. Olde Daalhuis, D. W. Lozier, B. I. Schneider, R. F. Boisvert, C. W. Clark, B. R. Miller, B. V. Saunders, H. S. Cohl, and M. A. McClain (eds.), \textit{NIST Digital Library of Mathematical Functions, Chap. 12: Parabolic Cylinder Functions} [\url{https://dlmf.nist.gov/12}].
\bibitem{Olver1959} F. W. J. Olver, J. Res. Natl. Bur. Stand. \textbf{63B}, 131 (1959).
\bibitem{Gundlach1966} K. H. Gundlach, Solid-State Electron. \textbf{9}, 949 (1966).
\bibitem{Li2020} W. Li, K. Nomoto, D. Jena, and H. G. Xing, Appl. Phys. Lett. \textbf{117}, 222104 (2020).
\bibitem{Forbes2006} R. G. Forbes, Appl. Phys. Lett. \textbf{89}, 113122 (2006).
\bibitem{Heine1963} V. Heine, Proc. Phys. Soc. \textbf{81}, 300 (1963).
\bibitem{Chang1982} Y.-C. Chang and J. N. Schulman, Phys. Rev. B \textbf{25}, 3975 (1982).
\bibitem{Kane1961} E. O. Kane, J. Appl. Phys. \textbf{32}, 83 (1961).
\bibitem{Hatakeyama2004} T. Hatakeyama, T. Watanabe, T. Shinohe, K. Kojima, K. Arai, and N. Sano, Appl. Phys. Lett. \textbf{85}, 1380 (2004).
\bibitem{Okino2017} H. Okino, N. Kameshiro, K. Konishi, A. Shima, and R. Yamada, J. Appl. Phys. \textbf{122}, 235704 (2017).
\bibitem{WernerGuttler1991} J. H. Werner and H. H. G\"uttler, J. Appl. Phys. \textbf{69}, 1522 (1991).
\bibitem{Tung1992} R. T. Tung, Phys. Rev. B \textbf{45}, 13509 (1992).
\bibitem{SentaurusUG} \textit{Sentaurus Device User Guide, Version W-2024.09} (Synopsys, Mountain View, CA, 2024).
\bibitem{Franz1956} W. Franz, in \textit{Handbuch der Physik}, ed. S. Fl{\"u}gge (Springer, Berlin, 1956) Vol.~17, p.~155.
\end{thebibliography}
\end{document}